\documentclass{aa}
\usepackage{graphicx}
\begin{document}
   \title{A study of the low-frequency quasi-periodic oscillations in the
   X-ray light curves of the black hole candidate \object{XTE~J1859+226}}

   \author{
           P. Casella\inst{1}\fnmsep\inst{2}\fnmsep\thanks{Present address:
           INAF - Osservatorio Astronomico di Brera, via E. Bianchi 46,
           I--23807 Merate (LC), Italy},
           T. Belloni\inst{3},
           J. Homan\inst{4}
           and L. Stella\inst{1}
          }

   \offprints{casella@mporzio.astro.it}

   \institute{
              INAF - Osservatorio Astronomico di Roma,
              Via di Frascati, 33, I--00040 Monte Porzio Catone (Roma), Italy\\
         \and
              Physics Department, Universit\`a degli Studi ``Roma Tre'',
              Via della Vasca Navale 84, I--00146 Roma, Italy\\
         \and
              INAF - Osservatorio Astronomico di Brera,
              via E. Bianchi 46, I--23807 Merate (LC), Italy\\
         \and
              Center for Space Researh, Massachusetts Institute of Technology,
              77 Massachusetts Avenue, Cambridge, MA 02139, USA\\
             }

     \date{Received May 06, 2004; accepted July 06, 2004}

   \abstract{
   We present the results of an extensive timing analysis
   of the 1999 outburst of the soft X-ray transient and black hole
   candidate \object{XTE~J1859+226} as observed with the {\it Rossi X-Ray
   Timing Explorer}. Three main different types of low frequency (1-9 Hz)
   quasi-periodic oscillations (QPOs) were observed and classified,
   strenghtening the general picture that is emerging for the variability
   properties of black hole X-ray binaires. Rapid transitions between
   different power spectral shapes were observed and their link with the count
   rate was studied. Furthermore, we show that a frequency of $\sim$6 Hz seems
   to hold a particular place: one of the three QPO types we found was very
   stable when at this frequency, as it happens
   in Z sources as well. The coherence of its subharmonic peak was
   higher when the fundamental was close to 6 Hz, thus suggesting the
   presence of some resonance at this frequency. 
   }
   \titlerunning{A variability study of the Black Hole Candidate
   XTE~J1859+226}
   \authorrunning{P.Casella et al.}
   \maketitle


\section{Introduction}

Observations with the  {\it Rossi X-ray Timing Explorer (RXTE)} have
led to an extraordinary progress in the knowledge of the variability
properties of black-hole candidates (BHCs) in X-ray binaries (see
e.g. \cite{vanderKlis00} and \cite{Remillardetal02a}).  The fast
quasi-periodic oscillations (QPOs) that were discovered in many of
these systems are thought to originate in the innermost regions of
the accretion flows around stellar-mass black holes. Even though the
mechanism responsible for the QPOs is still unknown, the study of
their properties and behaviour can provide important clues on the
physics of accretion onto BHCs.

   \begin{figure*}[t]
     \centering
     \includegraphics[width=17.5cm]{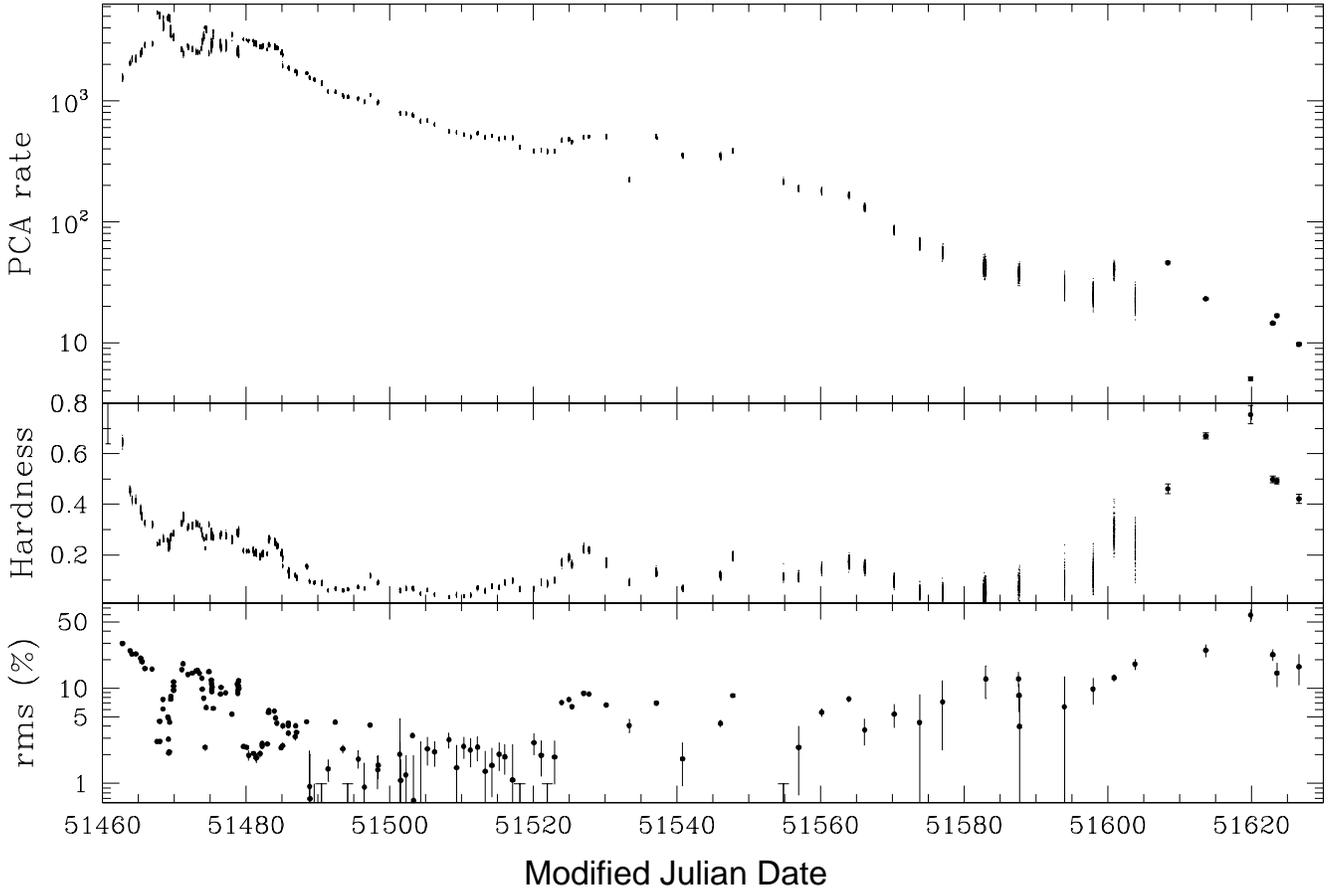}
     \caption{RXTE/PCA light curve (PCUs 0 and 2, {\it top panel}), hardness
    ratio [(7-15 keV)/(2-7 keV),{\it middle panel}] and total rms (0.03-64 Hz,
    {\it bottom panel}) of \object{XTE~J1859+226} during its 1999 outburst. 
    The time resolution for the light- and hardness curves is 16 s except for
    the last five points, which have one point for each observation.}
   \label{outburst}
   \end{figure*}


While only a few BHCs show high-frequency QPOs (50-450 Hz, for a
recent review see \cite{McC&R04} and references therein),
low-frequency QPOs with frequencies ranging from a few mHz to
$\sim$10 Hz are a common feature. The low-frequency QPOs were already
known before the RXTE era (see \cite{vanderKlis95} for an overview).
While only a few detections were reported on the basis of EXOSAT data
in the early 80$'$s, the Ginga satellite showed clear QPO features in
the power density spectra (PDS) of BHCs (see \cite{vanderKlis95}).
With RXTE, low-frequency QPOs were detected in virtually all observed
BHCs (see \cite{vanderKlis04}).  Usually these QPOs are associated
with the spectrally hard and intermediate states (\cite{vanderKlis95},
\cite{McC&R04}): they appear
together with a flat-top noise component, and on a time scale of days
their frequency often correlates with the source count rate (see e.g.
\cite{Cui98} and \cite{Reigetal00}). However, in the Ginga
observations of the bright transient \object{GS~1124--683} two distinct types
of low frequency QPOs were identified: one associated with a flat-top
noise component and one to a steep noise component
(\cite{Takizawaetal97}). The first type showed a strong dependence on
count rate, while the other had a rather stable frequency. These two
QPOs had centroid frequencies in the 1-10 Hz range and were clearly
related to the two PDS `flavors' of very high state observed in this
system (\cite{Miyamotoetal93}).

Wijnands et al. (1999) and Homan et al. (2001) reported on two
different types of QPOs in the RXTE data of \object{XTE~J1550--564}: a broad
one (type-A), with a quality factor Q (the QPO frequency divided by
the QPO full-width-half-maximum (FWHM)) of less than 3, and a
narrower one (type-B), with a Q larger than 6. Both QPOs were
characterized by a centroid frequency of 6 Hz and associated with a
weak red-noise component, but with different phase-lag behaviours.
\object{XTE~J1550--564} also showed the more common QPO-type associated with a
flat-top noise component (see \cite{Cuietal99} and
\cite{Sobczaketal00}). \cite{Remillardetal02b} dubbed this QPO
`type-C': its features are a high coherence (Q $\ga$ 10), a variable
centroid frequency (in the range 0.1 - 10 Hz) and a strong flat-top
noise component ($\sim$10--40\% rms) (see Table \ref{qpotable}).

While type-C QPOs are observed in many systems, the other two are
less common. In addition to \object{XTE~J1550--564}, type-B QPOs were also
observed in \object{GX~339--4} (e.g. \cite{Miyamotoetal91} and
\cite{Nespolietal03}),  \object{GRS~1739--278} (\cite{Wijnandsetal01}) and
possibly in \object{4U~1630--47} (\cite{TomsickKaaret00}), while type-A QPOs
were observed in \object{GX~339--4} (\cite{Nespolietal03}) and possibly in
\object{4U~1630--47} (\cite{TomsickKaaret00} and \cite{Dietersetal00}). 
Furthermore, in the light of the A-B-C classification, the two QPOs
observed in \object{GS~1124--683} (see above) can be tentatively identified
with types B and C, although a detailed analysis of Ginga data is
necessary to confirm this association.


   \begin{table}
\scriptsize
      \caption[]Summary of type-A, -B and -C QPOs properties in \object{XTE
      J1550--564} (\cite{Wijnandsetal99}, \cite{Homanetal01} and
      \cite{Remillardetal02b})
         \label{qpotable}
     $$
         \begin{array}{lccc}
            \hline
            \noalign{\smallskip}
            Property & Type A & Type B & Type C \\
            \noalign{\smallskip}
	    \hline
            \noalign{\smallskip}
            Frequency~(Hz) & \sim6 & \sim6 & 0.1-10 \\
	    Q~(\nu/FWHM) & \la3 & \ga6 & \ga10 \\
	    Amplitude~(\%rms) & 3-4 & \sim4 & 3-16 \\
	    Noise & ~~$weak$~$red$~~ & ~~$weak$~$red$~~ & ~~$strong$~$flat-top$~~ \\
            \noalign{\smallskip}
            \hline
         \end{array}
     $$
   \end{table}


   \begin{table*}
\scriptsize
      \caption[]{Power-spectral classification and variability parameters$^
{\mathrm{a}}$.}
         \label{parameters}
     $$
         \begin{array}{ccccccccc}
            \hline
            \noalign{\smallskip}
            Obs.~Id. & Date & MJD & ~QPO~Type~ & \nu_{qpo} &
            ~~FWHM_{qpo}~~~ & rms_{qpo}^{\mathrm{b}} & ~~Phase
            Lag^{\mathrm{c}} & 0.03$-$64~Hz~rms^{\mathrm{b}} \\
            \noalign{\smallskip}
	     & & & & (Hz) & (Hz) & \% & (rad) & (\%) \\
            \noalign{\smallskip}
	    \hline
            \noalign{\smallskip}
            $40124-01-04-00$ & ~1999~Oct~11~ & ~51462.768~ & C & 1.2 & 0.18
            & 16.61 & ~~0.00 $$\pm$$ 0.01 & 29.8 $$\pm$$ 0.3\\
            $40124-01-05-00$ & 1999~Oct~12 & 51463.833 & C & 3.05 & 0.31
            & 13.48 & -0.04 $$\pm$$ 0.01 & 24.9 $$\pm$$ 0.2 \\
            $40124-01-06-00$ & 1999~Oct~13 & 51464.109 & C & 3.64 & 0.46
            & 13.43 & -0.07 $$\pm$$ 0.01 & 22.9 $$\pm$$ 0.1 \\
            $40124-01-07-00$ & `` & 51464.633 & C & 3.65 & 0.46
            & 12.97 & -0.04 $$\pm$$ 0.01 & 23.1 $$\pm$$ 0.2 \\
            $40124-01-08-00$ & 1999~Oct~14 & 51465.305 & C & 4.39 & 0.48
            & 11.85 & -0.08 $$\pm$$ 0.01 & 20.78 $$\pm$$ 0.08 \\
            $40124-01-09-00$ & `` & 51465.498 & C & 4.95 & 0.56
            & 10.85 & -0.09 $$\pm$$ 0.01 & 19.16 $$\pm$$ 0.08 \\
            $40124-01-10-00$ & `` & 51465.902 & C & 5.81 & 0.72
            & 8.52 & -0.14 $$\pm$$ 0.01 & 16.2 $$\pm$$ 0.1 \\
            $40124-01-11-00$ & 1999~Oct~15 & 51466.896 & C & 5.97 & 0.65
            & 8.11 & -0.18 $$\pm$$ 0.01 & 16.0 $$\pm$$ 0.08 \\
            $40124-01-12-00$ & 1999~Oct~16 & 51467.581 & A & 7.75 & 3.85
            & 1.75 & -0.33 $$\pm$$ 0.07 & 2.76 $$\pm$$ 0.05 \\
            $40124-01-13-00$ & `` & 51467.961 & B & 6.08 & 0.45
            & 2.10 & ~~0.07 $$\pm$$ 0.07 & 4.5 $$\pm$$ 0.06 \\
            $40124-01-13-00$ & `` & 51467.973 & A & 8.42 & 2.885
            & 1.62 & -0.3 $$\pm$$ 0.1 & 2.8 $$\pm$$ 0.1 \\
            $40124-01-14-00$ & 1999~Oct~17 & 51468.427 & C^{*} & 8.64 & 2.37
            & 2.83 & -0.56 $$\pm$$ 0.04 & 7.67 $$\pm$$ 0.04 \\
            $40124-01-14-00$ & `` & 51468.427 & B & 6.43 & 0.82
            & 3.13 & ~~0.3 $$\pm$$ 0.09 & 6.1 $$\pm$$ 0.1 \\
            $40122-01-01-03$ & 1999~Oct~18 & 51469.093 & B & 5.96 & 0.75
            & 3.04 & ~~0.23 $$\pm$$ 0.04 & 5.02 $$\pm$$ 0.04 \\
            $40122-01-01-02$ & `` & 51469.159 & B & 5.96 & 0.85
            & 2.52 & ~~0.13 $$\pm$$ 0.04 & 4.83 $$\pm$$ 0.05 \\
            $40122-01-01-02$ & `` & 51469.179 & A & 7.85 & 3.26
            & 1.79 & -0.02 $$\pm$$ 0.1 & 2.9 $$\pm$$ 0.1 \\
            $40122-01-01-01$ & `` & 51469.226 & A & 7.74 & 5.22
            & 1.37 & -0.84 $$\pm$$ 0.2 & 2.1 $$\pm$$ 0.1 \\
            $40122-01-01-00$ & `` & 51469.293 & A & 7.59 & 3.81
            & 1.64 & -0.33 $$\pm$$ 0.1 & 2.16 $$\pm$$ 0.09 \\
            $40122-01-01-00$ & `` & 51469.360 & B & 5.94 & 0.80
            & 2.02 & ~~0.125 $$\pm$$ 0.06 & 4.40 $$\pm$$ 0.06 \\
            $40124-01-16-00$ & `` & 51469.493 & C^{*} & 8.45 & 3.54
            & 3.10 & -0.49 $$\pm$$ 0.04 & 7.69 $$\pm$$ 0.07 \\
            $40124-01-16-00$ & `` & 51469.517 & C^{*} & 7.7 & 1.36
            & 4.53 & -0.29 $$\pm$$ 0.04 & 8.28 $$\pm$$ 0.09 \\
            $40124-01-17-00$ & `` & 51469.893 & C^{*} & 7.89 & 1.17
            & 3.22 & -0.41 $$\pm$$ 0.07 & 9.6 $$\pm$$ 0.1 \\
            $40124-01-17-00$ & `` & 51469.899 & C^{*} & 7.54 & 1.58
            & 5.07 & -0.26 $$\pm$$ 0.04 & 10.6 $$\pm$$ 0.1 \\
            $40124-01-17-00$ & `` & 51469.905 & C^{*} & 7.27 & 1.19
            & 5.46 & -0.22 $$\pm$$ 0.04 & 11.7 $$\pm$$ 0.1 \\
            $40124-01-18-00$ & 1999~Oct~20 & 51471.024 & C & 5.88 & 0.42
            & 6.98 & -0.17 $$\pm$$ 0.015 & 15.81 $$\pm$$ 0.07 \\
            $40124-01-19-00$ & `` & 51471.224 & C & 5.18 & 0.61
            & 9.10 & -0.15 $$\pm$$ 0.015 & 18.11 $$\pm$$ 0.07 \\
            $40124-01-20-00$ & `` & 51471.890 & C & 6.47 & 0.80
            & 6.59 & -0.25 $$\pm$$ 0.025 & 14.04 $$\pm$$ 0.07 \\
            $40124-01-21-00$ & 1999~Oct~21 & 51472.503 & C & 6.34 & 0.90
            & 7.29 & -0.2 $$\pm$$ 0.02 & 14.49 $$\pm$$ 0.06 \\
            $40124-01-15-00$ & 1999~Oct~22 & 51473.245 & C & 6.13 & 0.72
            & 7.11 & -0.22 $$\pm$$ 0.04 & 15.5 $$\pm$$ 0.1 \\
            $40124-01-23-00$ & `` & 51473.822 & C^{*} & 6.92 & 0.93
            & 5.68 & -0.27 $$\pm$$ 0.03 & 12.79 $$\pm$$ 0.09 \\
            $40124-01-23-01$ & `` & 51473.890 & C^{*} & 7.73 & 1.59
            & 4.00 & -0.43 $$\pm$$ 0.03 & 9.85 $$\pm$$ 0.07 \\
            $40124-01-15-02$ & 1999~Oct~23 & 51474.088 & C^{*} & 7.39 & 4.52
            & 3.63 & -0.43 $$\pm$$ 0.03 & 7.91 $$\pm$$ 0.08 \\
            $40124-01-24-00$ & `` & 51474.429 & B-Cath. & 5.84 & 0.81
            & 3.62 & ~~0.23 $$\pm$$ 0.04 & 6.27 $$\pm$$ 0.06 \\
            $40124-01-25-00$ & `` & 51474.820 & C & 6.19 & 0.68
            & 7.18 & -0.22 $$\pm$$ 0.02 & 14.9 $$\pm$$ 0.1 \\
            $40124-01-26-00$ & 1999~Oct~24 & 51475.154 & C^{*} & 7.06 & 0.95
            & 5.06 & -0.29 $$\pm$$ 0.03 & 12.28 $$\pm$$ 0.08 \\
            $40124-01-26-00$ & `` & 51475.166 & C^{*} & 7.24 & 1.06
            & 4.43 & -0.34 $$\pm$$ 0.03 & 11.31 $$\pm$$ 0.07 \\
            $40124-01-26-00$ & `` & 51475.177 & C^{*} & 7.73 & 1.36
            & 3.54 & -0.40 $$\pm$$ 0.03 & 9.84 $$\pm$$ 0.07 \\
            $40124-01-26-00$ & `` & 51475.218 & C^{*} & 7.75 & 1.72
            & 3.70 & -0.45 $$\pm$$ 0.02 & 9.29 $$\pm$$ 0.04 \\
            $40124-01-26-00$ & `` & 51475.252 & C^{*} & 7.56 & 1.24
            & 3.65 & -0.33 $$\pm$$ 0.03 & 10.35 $$\pm$$ 0.07 \\
            $40124-01-27-00$ & `` & 51475.428 & B-Cath. & 5.79 & 0.84
            & 4.28 & ~~0.25 $$\pm$$ 0.07 & 6.2 $$\pm$$ 0.1 \\
            $40124-01-28-00$ & 1999~Oct~25 & 51476.428 & C^{*} & 7.64 & 2.1
            & 2.91 & -0.48 $$\pm$$ 0.03 &  8.69 $$\pm$$ 0.06 \\
            $40124-01-28-01$ & `` & 51476.501 & C^{*} & 7.52 & 1.53
            & 4.22 & -0.36 $$\pm$$ 0.04 & 10.27 $$\pm$$ 0.08 \\
            $40124-01-29-00$ & 1999~Oct~26 & 51477.152 & C^{*} & 7.54 & 2.4
            & 3.66 & -0.42 $$\pm$$ 0.02 & 8.99 $$\pm$$ 0.04 \\
            $40124-01-30-00$ & 1999~Oct~27 & 51478.017 & B & 5.06 & 0.7
            & 3.79 & ~~0.17 $$\pm$$ 0.04 & 5.36 $$\pm$$ 0.05 \\
            $40124-01-31-00$ & `` & 51478.777 & C^{*} & 7.26 & 0.96
            & 3.97 & -0.33 $$\pm$$ 0.03 & 11.12 $$\pm$$ 0.07 \\
            $40122-01-02-00$ & `` & 51478.816 & C^{*} & 7.77 & 1.31
            & 2.74 & -0.40 $$\pm$$ 0.05 & 9.51 $$\pm$$ 0.08 \\
            $40122-01-02-00$ & `` & 51478.844 & C^{*} & 7.77 & 1.89
            & 3.38 & -0.44 $$\pm$$ 0.05 & 8.9 $$\pm$$ 0.1 \\
            $40122-01-02-00$ & `` & 51478.882 & C^{*} & 7.38 & 1.25
            & 4.00 & -0.36 $$\pm$$ 0.04 & 10.72 $$\pm$$ 0.09 \\
            $40122-01-02-00$ & `` & 51478.913 & C^{*} & 7.54 & 1.39
            & 3.59 & -0.40 $$\pm$$ 0.04 & 10.45 $$\pm$$ 0.08 \\
            $40122-01-02-00$ & `` & 51478.949 & C^{*} & 7.04 & 0.96
            & 4.84 & -0.37 $$\pm$$ 0.03 & 11.99 $$\pm$$ 0.07 \\
            $40122-01-02-00$ & `` & 51478.980 & C^{*} & 7.46 & 1.03
            & 3.12 & -0.49 $$\pm$$ 0.05 & 10.01 $$\pm$$ 0.08 \\
            $40124-01-36-00$ & 1999~Nov~01 & 51483.117 & B & 4.69 & 0.63
            & 4.06 & ~~0.16 $$\pm$$ 0.03 & 5.58 $$\pm$$ 0.07 \\
            $40124-01-36-00$ & `` & 51483.158 & B & 4.69 & 0.54
            & 3.93 & ~~0.22 $$\pm$$ 0.04 & 5.88 $$\pm$$ 0.06 \\
            $40124-01-37-00$ & `` & 51483.945 & B & 4.59 & 0.54
            & 4.31 & ~~0.16 $$\pm$$ 0.03 & 5.80 $$\pm$$ 0.05 \\
            $40124-01-37-01$ & 1999~Nov~02 & 51484.077 & B & 4.53 & 0.62
            & 4.05 & ~~0.07 $$\pm$$ 0.05 & 4.89 $$\pm$$ 0.15 \\
            $40124-01-37-02$ & `` & 51484.276 & B & 4.45 & 0.61
            & 4.03 & ~~0.09 $$\pm$$ 0.03 & 4.33 $$\pm$$ 0.06 \\
            $40124-01-39-00$ & `` & 51485.875 & ? & 5.61 & 2.3
            & 2.68 & -0.51 $$\pm$$ 0.04 & 4.1 $$\pm$$ 0.1 \\
            $40124-01-40-00$ & 1999~Nov~04 & 51486.828 & ? & 5.37 & 1.7
            & 3.00 & -0.43 $$\pm$$ 0.07 & 3.1 $$\pm$$ 0.3 \\
            $40124-01-40-01$ & `` & 51486.873 & ? & 5.22 & 1.3
            & 2.89 & -0.3 $$\pm$$ 0.1 & 4.0 $$\pm$$ 0.2 \\
            $40124-01-41-00$ & 1999~Nov~05 & 51487.009 & ? & 4.74 & 1.3
            & 2.18 & -0.42 $$\pm$$ 0.05 & 3.5 $$\pm$$ 0.1 \\
            $40124-01-42-00$ & 1999~Nov~06 & 51488.409 & ? & 7.08 & 2.4
            & 1.75 & -0.88 $$\pm$$ 0.06 & 4.46 $$\pm$$ 0.08 \\
            $40124-01-49-01$ & 1999~Nov~15 & 51497.255 & ? & 5.21 & 1.9
            & 2.82 & -0.36 $$\pm$$ 0.07 & 4.1 $$\pm$$ 0.2 \\
            $40124-01-58-01$ & 1999~Dec~15 & 51527.004 & ? & 9.1 & 0.8
            & 2.17 & ~~0.0 $$\pm$$ 0.2 & 8.9 $$\pm$$ 0.2 \\
            $40124-01-59-00$ & 1999~Dec~18 & 51530.133 & ? & 8.15 & 2.1
            & 2.91 & ~~0.0 $$\pm$$ 0.2 & 6.71 $$\pm$$ 0.25 \\
            $40124-01-61-00$ & 2000~Jan~03 & 51546.036 & ? & 4.65 & 0.8
            & 2.3  & -0.17 $$\pm$$ 0.25 & 4.3 $$\pm$$ 0.4 \\
            \noalign{\smallskip}
            \hline
         \end{array}
     $$
\begin{list}{}{}
\item[$^{\mathrm{a}}$] Only observations with evidence for low frequency QPO
  are listed.
\item[$^{\mathrm{b}}$] Normalization according to \cite{Belloni90}
\item[$^{\mathrm{c}}$] Phase lag in radians between the 2-5 and 5-13 keV light
  curves and integrated over the frequency range $\nu_{qpo}\;\pm\;FWHM/2$
\end{list}
   \end{table*}


These oscillations, whose nature is still not understood, provide a 
direct way to explore the accretion flow around black holes (and neutron 
stars). In particular, their association with specific spectral states
and the phenomenology that is emerging indicate that they are a key 
ingredient in understanding the physical conditions that give origin to
the different states.

The soft X-ray transient \object{XTE~J1859+226} was discovered on 1999 October
9 with the RXTE All Sky Monitor (\cite{Woodetal99}), which detected a
2--12 X-ray flux of $\sim$160 mCrab ($\sim 4.5 \times 10^{-9}$ erg
s$^{-1}$ cm$^{-2}$) quickly rising at a rate of $\sim$6 mCrab/hour.
A follow-up observation of the source with the RXTE/PCA (Proportional
Counter Array, 2--60 keV) revealed a hard power-law dominated
spectrum (\cite{Markwardtetal99}).  On October 16, the source reached
its peak flux of $\sim 5 \times 10^{-8}$ erg s$^{-1}$ cm$^{-2}$ in
the 2-80 keV band (corresponding to a luminosity of $\sim 7 \times
10^{38}$ erg s$^{-1}$ for an assumed distance of 11 kpc (Zurita et
al. 2002)).  After the initial hard rise, \object{XTE~J1859+226} softened
at its peak intensity and continued to soften for almost two months,
when a secondary hard plateau took place (see Fig. \ref{outburst}, MJD$\sim$
51520--51540).
This was rather similar to the behaviour observed in many X-ray
transients, in particular \object{XTE~J1550--564} (see \cite{Remillardetal02b}
and references therein), suggesting a common scenario for the evolution of
these objects. Both low-frequency ($\sim 1-4$ Hz and $\sim 6$ Hz) and
high-frequency ($\sim 150-187$ Hz) QPOs have been reported (see
\cite{Cuietal00} and \cite{Fockeetal00}) from \object{XTE~J1859+226}.

The optical (\cite{Garnavichetal99}) and radio
(\cite{PooleyHjellming99}) counterparts were identified soon after
the discovery of the source. Optical photometry revealed a possible
period of 9.15 hr (\cite{Garnavichetal00}).  Monitoring at radio
wavelengths suggested that at least two relativistic ejection
episodes took place approximately on October 16.5 (MJD = 51467.5) and 27 (MJD =
51478). However, no ejecta were spatially resolved at radio
wavelengths (\cite{Brocksoppetal02}).

Here we present the results of an extensive X-ray timing analysis of
the 1999 outburst of \object{XTE~J1859+226}, focussing on the low-frequency
QPOs. We found three different types of low frequency (1--10 Hz)
QPOs. We show that these correspond to the above-mentioned A, B and C
QPOs types, all of them showing distinctive and well defined
behaviours and phase lags .

   \begin{figure}[t]
     \centering
     \includegraphics[width=8.5cm]{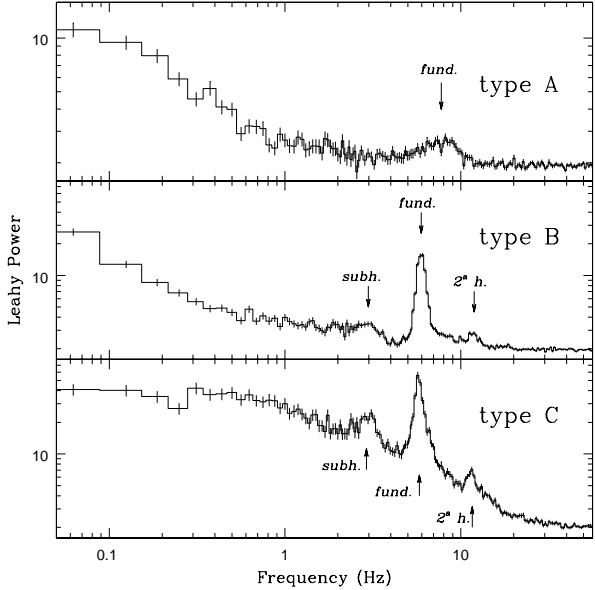}
     \caption{Examples of type A, B and C QPOs from our \object{XTE~J1859+226} 
observations. QPO and harmonics peaks are indicated.
       {\it Upper panel:} obs. 40124-01-12-00.
       {\it Middle panel:} obs. 40122-01-01-03.
       {\it Bottom panel:} obs. 40124-01-10-00).
       The Poisson noise was not subtracted.}
     \label{3types}
   \end{figure}



\section{Observations and data analysis}

We analyzed 129 RXTE/PCA observations made during the 1999 outburst
of the black-hole candidate \object{XTE~J1859+226}, between MJD 51462
(1999-10-11) and 51626 (2000-03-23). Table \ref{parameters} shows dates
and parameters of the observations where a low frequency QPO has been observed.

The PCA data were obtained in several simultaneous different modes
(see Tab. \ref{XTEmodes}). Only proportional counter units (PCUs) 0
and 2 were always active during our observations. {\tt Standard 2}
data from these two PCUs were used to create light- and hardness
curves for the whole outburst, whereas the high time resolution data
from all active PCUs (in a given obseravtion) were used for the
timing analysis. A hardness ratio was defined as the ratio of counts
in the range 7-15 keV (12-31 channels) to those in the range 2-7 keV
(0-11). Fast Fourier Transforms were made from 16s data intervals with
Nyquiest frequencies of 64 Hz ({\tt Binned} data) and 4096 Hz ({\tt
Single bit} data). The resulting PDS were averaged, rebinned
logarithmically, and the Poissonian noise, including the Very Large
Events (VLE) contribution (\cite{Zhang95}, \cite{Zhangetal95}), was
subtracted. The PDS were normalised to fractional squared rms,
following \cite{Belloni90}. PDS fitting was carried out by using the
standard Xspec fitting package by using a one-to-one energy-frequency
  conversion and a unit response. Following \cite{Bellonietal02}, we fitted
the 
noise components with two Lorentzian shapes, one zero-centered  and a second
one centered at a few Hz. The QPOs were fitted with one Lorentzian each too, 
only occasionally needing the addition of a Gaussian
component to better approximate the shape of the narrow peaks and to
reach values of reduced $\chi^2$ close to 1. For the observations
where the dynamical power spectra showed transitions between
different power spectral shapes (see below), we separated different
time intervals in order to obtain average power spectra for each
shape.


   \begin{table}
\scriptsize
      \caption[]{RXTE/PCA data modes active during the \object{XTE~J1859+226} 
observations}
         \label{XTEmodes}
     $$
         \begin{array}{llcc}
            \hline
            \noalign{\smallskip}
            Mode & Time~res. & Number~of & PHA \\
            Name & \hspace{2.0mm}(sec.) & PHA~Channels & Energy~range~(keV) \\
            \noalign{\smallskip}
            \hline
            \noalign{\smallskip}
            {\tt Standard1} & \hspace{4.0mm}2^{-3} & 1 & 2-60  \\
            {\tt Standard2} & \hspace{4.0mm}2^{~4} & 129 & 2-60  \\
            {\tt Binned}    & \hspace{4.0mm}2^{-7} & 36 & 2-15  \\
            {\tt SB1}       & \hspace{4.0mm}2^{-13} & 14 & 2-6  \\
            {\tt SB2}       & \hspace{4.0mm}2^{-13} & 22 & 6-15  \\
            {\tt Event}     & \hspace{4.0mm}2^{-16} & 16 & 15-60  \\
            \noalign{\smallskip}
            \hline
         \end{array}
     $$
   \end{table}


   \begin{figure}[t]
     \centering
     \includegraphics[width=8.5cm]{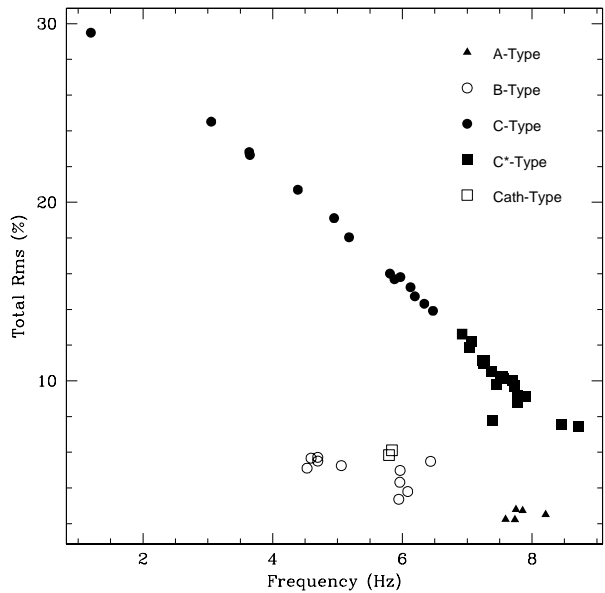}
     \caption{Centroid QPO frequency vs. 0.03-64 Hz fractional rms of the
     detected QPOs. Each point corresponds to a different observation, except
     for the few cases in which transitions between different pds shapes have
     been observed. Error bars are smaller than symbols.}
     \label{freq-rms}
   \end{figure}

For every 16 s interval we also produced a cross-spectrum between the
2-5 and 5-13 keV resolved light curves (defined as $C(j)=X_1^*(j) X_2(j)$,
where $X_1$ and $X_2$ are the complex Fourier coefficients for the two energy
bands at a frequency  $\nu_j$), calculated average
cross-spectrum vectors for each observation, and then derived a phase
lag as a function of frequency from the angle in the complex plane of
these vectors ($\phi_j=arg[C(j)]$). The error in $\phi$ is computed from the
observed variance of $C$ in the real and imaginary directions.
 In line with recent literature we defined phase lags
as positive when the hard X-ray variability {\it follows} the soft
one. To quantify the phase-lag behaviour of the QPOs, we extracted
the phase lags in a range centered at the QPO peak frequency and
corresponding to the width of the peak itself
($\nu_{p}\;\pm\;${\small FWHM/2}, see \cite{Reigetal00}). In Table
\ref{parameters}, we list the frequency, full-width half-maximum
(FWHM), 2-15 keV fractional rms, and phase lag of
the QPO for each observation in which one was detected. The total
integrated fractional rms (2-15 keV) of the PDS is given as well.

In Figure \ref{outburst}, we show the 2-60 keV  light curve of the whole
outburst, the hardness ratio and the integrated 0.03-64 Hz fractional rms of
the 2-15 keV light curves.

\section{Power spectra and phase lags}

In many of our observations QPOs were detected, with frequencies
ranging from $\sim$1 to $\sim$9 Hz. Three main types could be
distinguished, which, based on their phase lag and coherence
properties, could be associated with type-A, -B, and -C QPOs. Example
power spectra of each type are shown in Figure \ref{3types}. In
addition to these three main types we also identified various
sub-types, which will be discussed below.

   \begin{figure}[t]
     \centering
     \includegraphics[width=8.5cm]{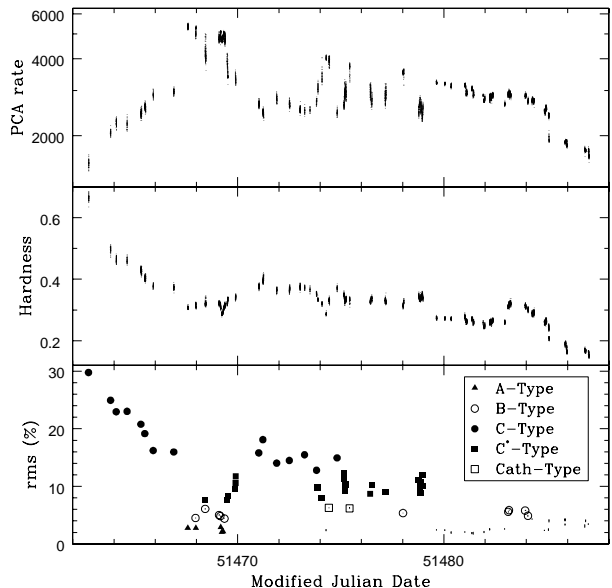}
     \caption{Same as Figure \ref{outburst}, but for only the
       first 25 days of
       the outburst. See inset for the symbols that are used to represent
       different QPO types in the lower panel - dots are used if no QPO was
       detected.}
     \label{zoom}
   \end{figure}

A useful method for differentiating between the three types of PDS is
shown in Figure \ref{freq-rms}, in which we plot the integrated
fractional rms of each PDS versus the centroid frequency of the QPO.
Several groups of points can be identified. The first large group of
points in the plot, type-C and C$^{*}$ (see paragraphs \$3.1 and
\$3.2), is diagonally spread across the plot, covering the whole
frequency range between 1 and 10 Hz and a large range in rms (7--30\%
rms). Another group, type-A (see paragraphs \$3.3), is clustered
around a frequency of 8 Hz and rms of 2\%. Finally a third group is
located at a slightly higher rms (4--6\%) in the 4.5 - 6.5 Hz range:
type-B and "B-Cathedral" (see paragraphs \$3.4 and \$3.5). A more
detailed analysis of the correlation between the frequency these
three QPO types and the integrated fractional rms in different
sources is described in a forthcoming paper (Casella et al. 2004, in
prep.). In Figure \ref{zoom} we show the light curve, the hardness
and the total fractional rms of the first 25 days of the outburst
(see Fig \ref{outburst} for energy and frequency ranges) and indicate
where the three types of QPO appear. In the bottom panel we have marked
the three types of QPO and their associated sub-types. In the following
paragraphs we describe all these types in detail, characterizing
their PDS and phase-lag behaviour.

\subsection{Type-C QPOs}

   \begin{figure}[t]
     \centering
     \includegraphics[width=8.5cm]{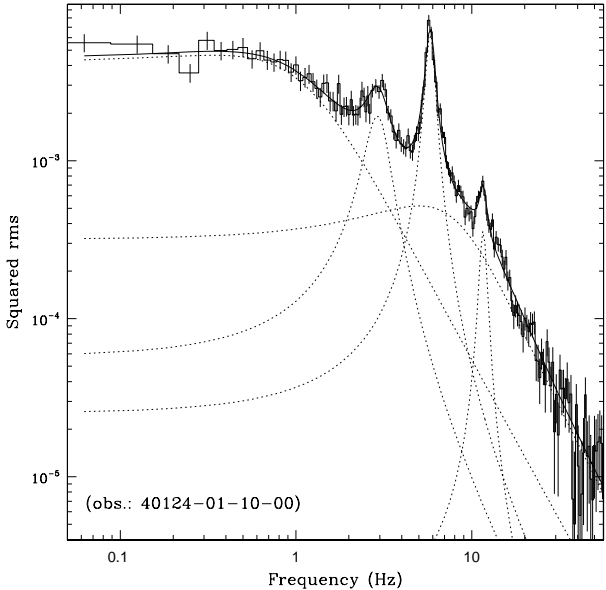}
     \caption{An example of type-C power spectrum (2--15 keV; obs.:
     40124-01-10-00). The solid line shows the best fit with five Lorentzians
     (dotted lines). See the correspondent lags in Fig. \ref{C_Lags}, fourth
     panel.}
     \label{typeC}
   \end{figure}

   \begin{figure}[t]
     \centering
     \includegraphics[width=9.0cm]{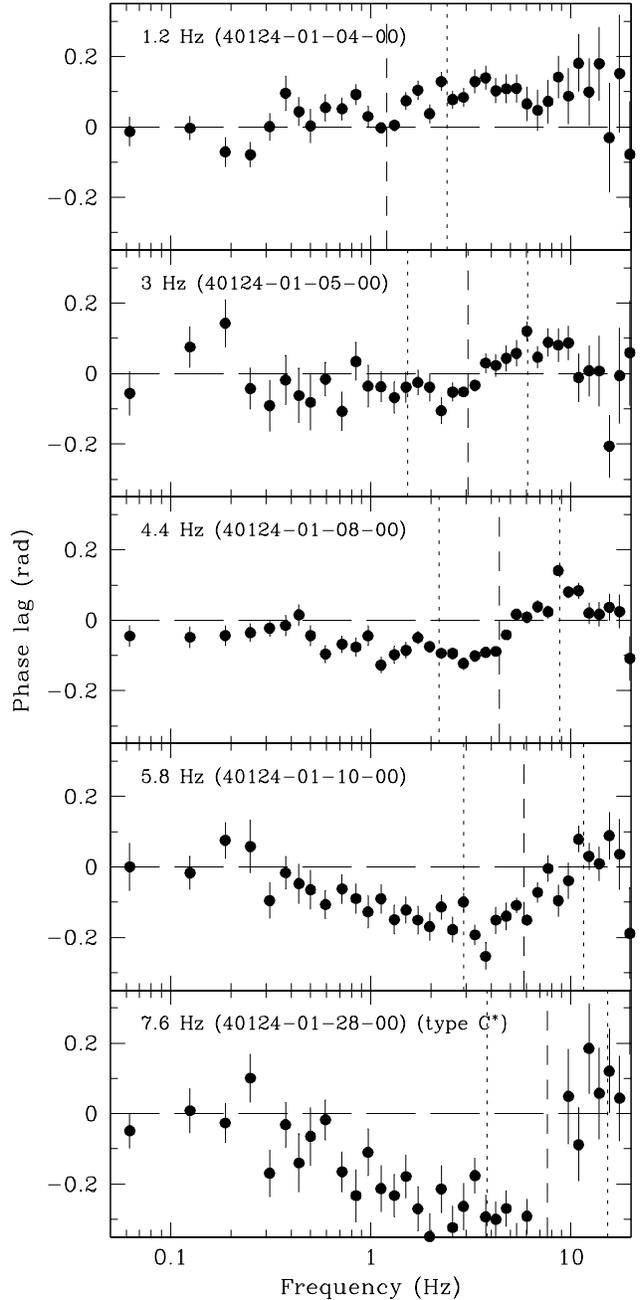}
     \caption{Type-C phase lags vs. frequency for four different QPO
       frequencies. Bottom panel: an example of type-C$^{*}$ phase
       lags. Positive values indicate that the hard (5--13 keV) photons are
       lagging the soft(2--5keV) photons. QPO centroid frequency and
       observation I.D. are indicated for each panel. The dashed lines mark
       the frequency of the QPO, while the dotted lines mark the subharmonic
       (if present) and second harmonic frequencies.}
     \label{C_Lags}
   \end{figure}

   \begin{figure}[t]
     \centering
     \includegraphics[width=8.5cm]{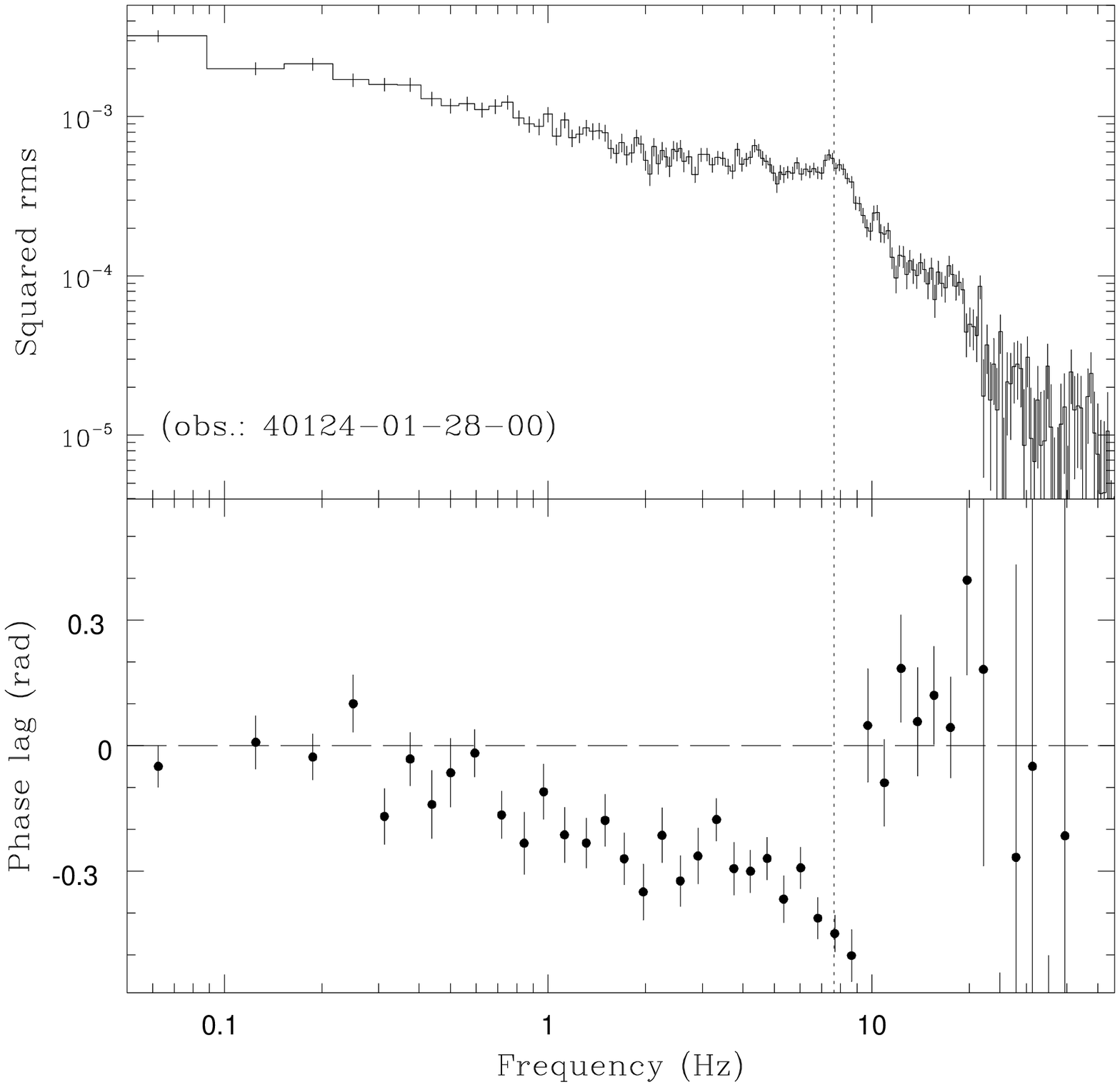}
     \caption{An example of type C$^{*}$ PDS (2--15 keV) and phase lags (obs.:
     40124-01-28-00). The dotted vertical line marks the frequency of the QPO.}
     \label{typeC*}
   \end{figure}

In the early stages of the outburst (see Table \ref{parameters} for
dates), when the source flux was rising quickly, the PDS showed four
main components: a strong (15-30\% rms) flat-topped noise and three 
harmonically related QPOs (see Fig \ref{typeC}). The central 1--7 Hz QPO was 
strong (rms amplitude 6--16\%) and narrow (Q $\sim$7--10). Note
that in a few cases the addition of a Gaussian component was required, in
order to better approximate the peak's shape. This was also the case for the
second harmonic, which was always present. A broad (Q
$\sim$2) subharmonic was also detected in all cases, except for the
first observation (40124-01-04-00) where the frequency of the
fundamental was at its lowest value). The harmonic relation
between the three peaks was confirmed by allowing the centroids to vary
independently. We decided to fix the centroid frequencies to
harmonic ratios to obtain self-consistent estimates of the amplitude
and width.

   \begin{figure}[t]
     \centering
     \includegraphics[width=8.5cm]{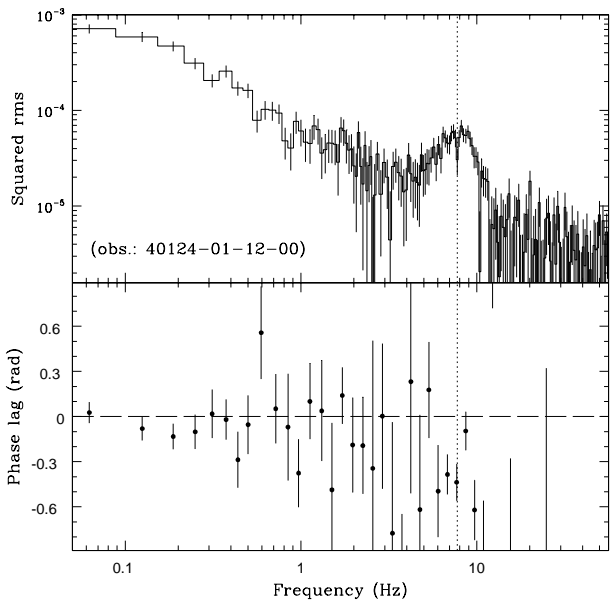}
     \caption{An example of type-A
     PDS (2--15 keV) and phase lags (obs.: 40124-01-12-00). The dotted
     vertical line marks the frequency of the QPO. Lags at frequencies above
     $\sim 10$ Hz are, given the lack of statistics, random scattering between
     $-\pi$ and $\pi$.}
     \label{typeA}
   \end{figure}


The phase-lag behaviour of the type-C power spectra is similar to
that found in \object{GRS~1915+105} by Reig et al. (2000). It is strongly
correlated
with the frequency of the QPO, with a trend towards negative lags for
increasing QPO frequency. Figure \ref{C_Lags} (all panels except the
bottom one) shows four examples covering the whole observed range in
QPO frequency. Owing to poor statistics, lags became unmeasurable at
high frequencies; we thus plot them only below 20 Hz. In all observed
cases, the fundamental of the QPO showed negative lags (see Figure
\ref{rms_lags} and the Discussion), with a clear
trend towards zero for decreasing centroid frequency, consistent with
the Reig et al. (2000) results on \object{GRS~1915+105} (where peak lags are
positive at frequencies below $\sim$1 Hz). The lags of subharmonic
peak were always negative as well, while the second harmonic always
showed positive lags. It is important to mention the absence in the
cross spectra of narrow features standing out at the frequency of the
QPO and harmonics peaks. This suggests that we are measuring the
phase-lags of the underlying noise continuum rather than the phase
lags associated with the QPO peaks. 

\subsection{Type-C$^{*}$ QPOs}

Later in the outburst, type-C QPOs appeared again (see Table
\ref{parameters}). In this case, however, the centroid frequency was
higher and the rms amplitude lower. We refer to these QPOs as a
type-C$^{*}$. The power spectra showed a strong red-noise component and
a broad QPO peak at a centroid frequency of 7-9 Hz (Figure \ref{typeC*}, upper 
panel). Again a Gaussian component was added in many cases in order to better
approximate the peak shape. A second harmonic peak was sometimes
present, as well a subharmonic one. Phase lags were negative and large, up to
10 ms, over the range $\nu_{p}\;\pm\;${\small FWHM/2}, and decreased rapidly
at frequencies slightly higher than the QPO centroid, as can be
seen in Figure \ref{typeC*} (bottom panel).

Even though the QPO centroid frequency range was different from that
of type-C QPOs, and the rms and Q-values were smaller, the results
above (see also Figure \ref{freq-rms}) provide evidence that the
properties of
type-C and type-C$^{*}$ QPOs are smoothly connected if ordered for
increasing QPO frequency. The phase-lag behaviour confirms this: even though
the shape of a ``normal'' type-C QPO (Figure \ref{C_Lags}, upper panel) was
rather different from that of type C$^{*}$ QPOs (Figure \ref{C_Lags}, bottom
panel), a continuos transition between the two took place for
increasing QPO frequency. However, owing to the long time scale of the
QPO frequency variability (not detectable during one single RXTE observation)
it was not possible to observe a direct transition between the two QPO types.

\subsection{Type-A QPOs}

Type-A power spectra (see an example in Fig. \ref{typeA}, upper
panel) appeared at the peak of the outburst (see Table
\ref{parameters}), when the count rate was very high. They were
characterized by a broad QPO (Q $\sim$ 2), with centroid frequencies between 
7.5 and 8.5 Hz and fractional rms around $\sim$1.5\%, and a low amplitude (few
\% rms) red-noise component. Neither a
subharmonic nor a second harmonic was present. This was the QPO type
with the lowest total rms in our sample. The phase lags, except for a negative
excess around the QPO frequency, were consistent with zero.

\subsection{Type-B QPOs}

   \begin{figure}[t]
     \centering
     \includegraphics[width=8.5cm]{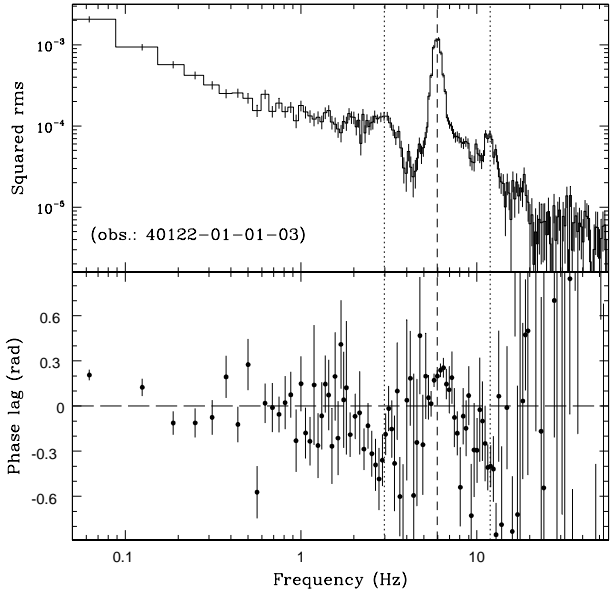}
     \caption{An example of type-B PDS (2--15 keV) and phase lags (obs.:
     40122-01-01-03). The dashed vertical line marks the frequency of the QPO,
     while the dotted lines mark the subharmonic and second harmonic
     frequencies.}
     \label{typeB}
   \end{figure}

The narrow (Q $\sim$10) QPO characterizing this type of PDS appeared
only in a rather restricted frequency range between 4.5 and 6.5 Hz
(see an example in Fig. \ref{typeB}, upper panel). The red-noise was
very weak (few \%  rms). The QPO peak profile was often more similar to a
Gaussian than a Lorentzian, but we needed to combine both components
in order to obtain a good fit (it is worth noticing that a similar
combination was used by Homan et al. (2001) and Wijnands et al.
(1999) for the \object{XTE~J1550--564} data, and by Nespoli et al. (2003) for
the \object{GX~339--4} data. In the latter case the authors explained the
Gaussian shape with the presence of centroid frequency variability on
a $\sim$10 s time scale. However, such variability was not detected
in the \object{XTE~J1859+226} data). A weak second harmonic was always
present, while a subharmonic appeared (with rms amplitudes $<$1 \%)
only when the fundamental frequency was at its highest values and the
fundamental and the second harmonic had low amplitudes. This
behaviour suggests the presence of a sort of ``balance'' between the
amplitude of the three peaks, particularly between the second
harmonic and subharmonic: when the subharmonic peak was at its
highest rms the second harmonic is not present, and viceversa.
Unfortunately, poor statistics did not allow a more precise
assessment of this. The lags (see Fig. \ref{typeB}, bottom panel)
were positive at the frequency of the fundamental peak (see Fig.
\ref{lags_fund_harm} for values) and negative at the frequencies of
subharmonic and second harmonic.

   \begin{figure}[t]
     \centering
     \includegraphics[width=8.5cm]{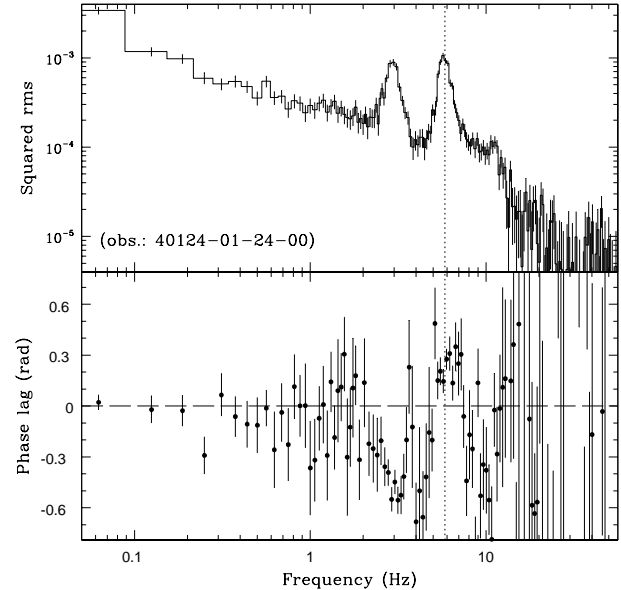}
     \caption{An example of type B-'Cathedral' PDS (2--15 keV) and phase lags
       (obs.: 40124-01-24-00). The dotted vertical line marks the frequency of
       the QPO.}
     \label{typeCath}
   \end{figure}

\subsection{Type B-'Cathedral' QPOs}

During two observations, 40124-01-24-00 (MJD: 51474.429) and
40124-01-27-00 (MJD: 51475.428), the power spectrum showed a peculiar
double-peaked QPO (see Fig \ref{typeCath}, upper panel), similar to
some of the type-B QPOs observed in \object{XTE~J1550--564}
(\cite{Wijnandsetal99}, \cite{Homanetal01}). In both observations,
two strong and narrow peaks were observed at harmonically related
frequencies of $\sim$3 and 6 Hz. In the case of the second observation we 
needed to add a Gaussian component to both peaks in order to better approximate
their shape. A weak red-noise component was present (the two Lorentzians
having  $\sim$2\% rms each). The rms amplitudes of the two QPO peaks were
$\sim$4\% and 2--2.5\%, for the 6 Hz and 3 Hz QPO, respectively. A weak
($\sim$1\% rms) peak at the harmonically related frequency of 12 Hz was
observed in both observations. The phase lags were consistent with zero over
the whole frequency range except at the frequencies where the QPOs were seen
(3 and 6 Hz). The lags of the 3 Hz QPO were negative and corresponded
(calculated, according with our definition, in the range
$\nu_{p}\;\pm\;${\small FWHM/2}) to a delay of $\sim$20 ms in the
{\it soft} X-ray variations, while for the $\sim$6 Hz peak the lags
were positive and corresponded to a delay of $\sim$7 ms in the {\it
hard} X-ray variations.

   \begin{figure}[t]
     \centering
     \includegraphics[width=8.5cm]{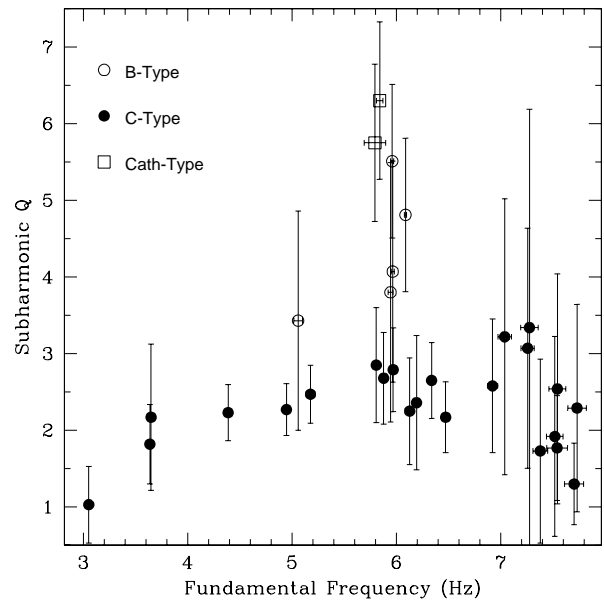}
     \caption{Coherence of the subharmonic peak for different pds types.}
     \label{subh}
   \end{figure}

   \begin{figure*}[t]
     \centering
     \includegraphics[width=18.0cm]{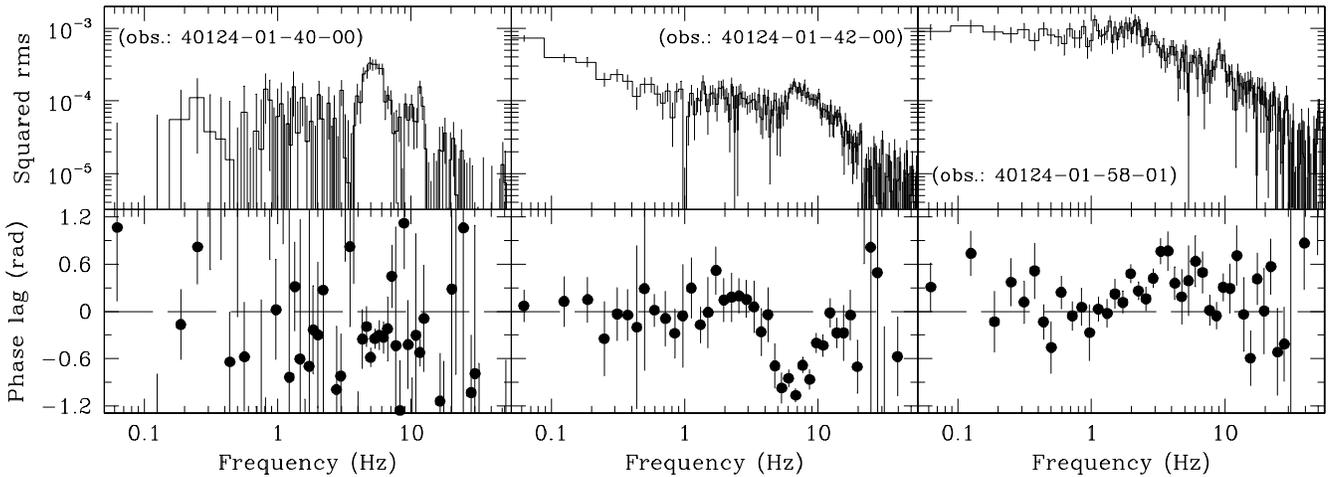}
     \caption{Three examples of unclassified power spectra (2--15keV) with
     their phase lags.}
     \label{spurie}
   \end{figure*}

   \begin{figure*}[t]
     \centering
     \includegraphics[width=17.5cm]{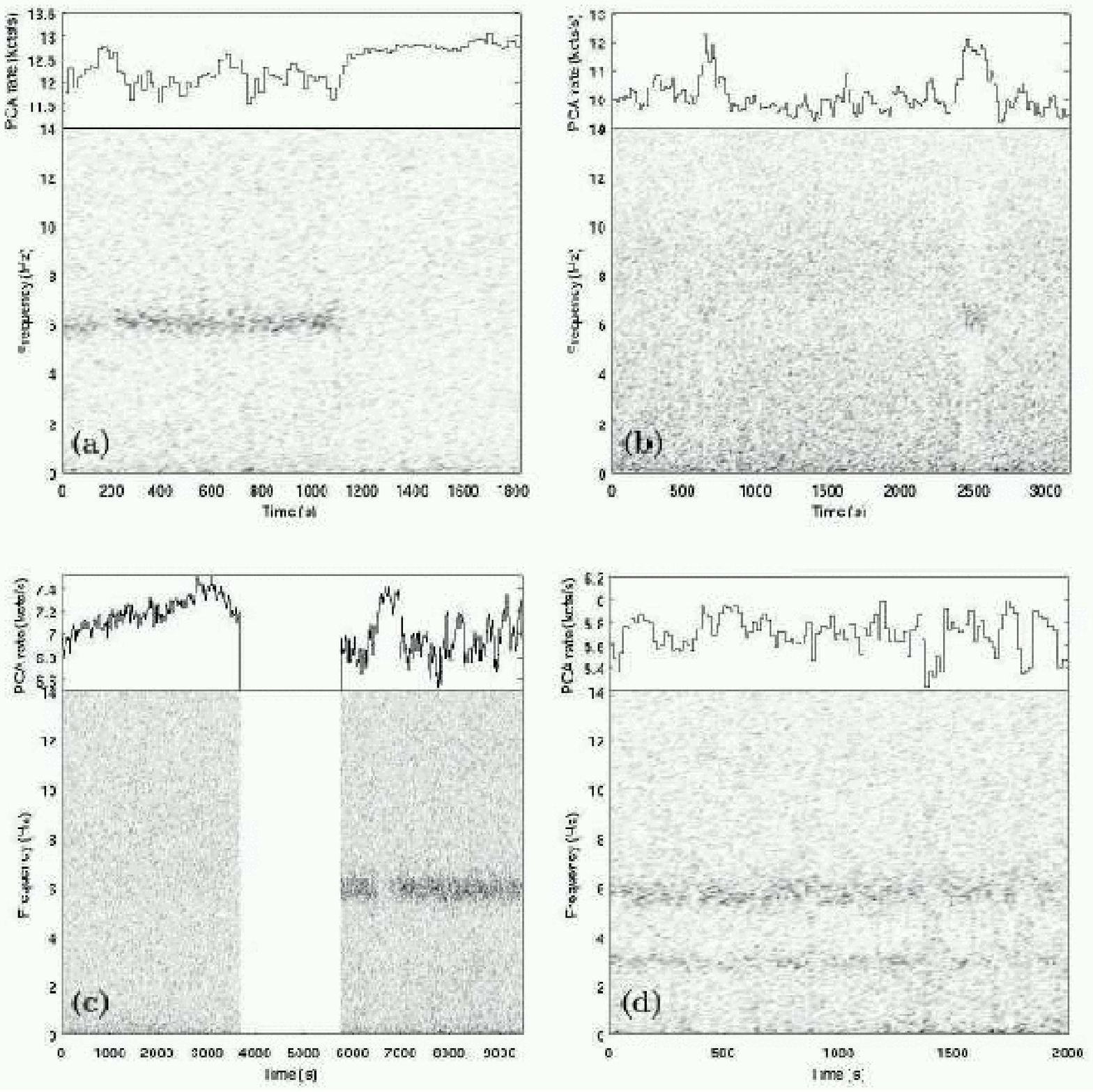}
     \caption{Light curves and dynamical PDS for observations 40124-01-13-00
       (panel (a), all five PCUs on), 40124-01-14-00 (panel (b), five PCUs
       on), 40122-01-01-00 (panel (c), three PCUs on) and 40124-01-24-00
       (panel (d), three PCUs on). The lower power value of the type-A QPO
       peaks with respect to the type-B render them invisible in the dynamical
       power spectra of panels (a) and (c). However, these are clearly seen in
       the total average PDS.}
     \label{trans}
   \end{figure*}


One might classify the $\sim$3 Hz peak as the fundamental and the
$\sim$6 Hz peak as the second harmonic. In this case, the $\sim$12 Hz
would be the fourth harmonic, while the third harmonic would be
missing or too weak to be detected. However, a comparison of this
``cathedral'' power spectrum with the others, suggests a similarity
with the type-B power spectrum. Apart from the high amplitude and
narrower width of the $\sim$3 Hz peak, the general characteristics
are almost identical: the total rms was around a few \% (compared to
$\sim$15\% for the type-C spectrum when the QPO was at $\sim$6 Hz),
and both the red-noise and the $\sim$6 Hz peak were weak. Moreover,
type-B PDS showed positive lags for the fundamental, as opposed to
those in the type-C power spectra. Thus it seems natural to identify
the $\sim$6 Hz peak (having positive lags) as the fundamental, the
$\sim$12 Hz as its second harmonic, and the $\sim$3 Hz peak as its
{\it subharmonic}. This is also in line with other PDS types in which
we identified subharmonic peaks. Moreover, the integrated rms of the
subharmonic peak of the B-cathedral PDS ($\sim 2-3 \%$) was not much
higher than that of other PDS. On the contrary, it was lower than that
of type-C QPOs (between 3 and 8 $\%$), and only slightly higher than
that of ``normal'' type-B QPOs ($\sim 1-2 \%$). However the coherence
of the B-cathedral subharmonic peak was the largest observed among
all low frequency QPOs (see Figure \ref{subh}), making the feature
more prominent. It is worth noticing that the $\sim$6 Hz QPO
frequency clearly stands out in figure \ref{subh} (see the
discussion).

\subsection{Special cases: unclassified QPOs}

In a few of the observations following the last type-B PDS (MJD
51484) we detected QPOs that we were not able to classify in terms of
the type A/B/C scheme. Owing to poor statistics and a too high
centroid frequency variability, we could not unambiguously fit the power
spectra and then obtain characteristic parameters. However, for sake
of completeness we report in Table \ref{parameters} the best estimate
for the parameters of each observation. In Figure \ref{spurie} we
show three representative examples of power spectra and phase lags
from these observations.

\section{Timing and spectral evolution}

In some observations, the dynamical PDS showed rapid (a few tens of
seconds) transitions between different power spectral shapes. In all
cases, the transitions involve type-B QPOs. In panels a-c of Figure
\ref{trans} we show three examples of different behaviours. In the
first half of the observation 40124-01-13-00 (MJD 51467.961, panel a), when
the light curve was highly variable, the PDS was of type-B (with a QPO
frequency $\sim 6$ Hz). Simultaneously with the rise observed in the light
curve after $\sim$1100 s from the start, the PDS showed a sharp transition
to a type-A shape with a QPO frequency of $\sim8$ Hz (not visible in
the gray scale representation). In the second part of the
observation, the light curve was much less variable and had a higher
mean count rate. Notice that a brief interval with the same
characteristics (type-A QPO, higher flux) was seen $\sim$200 s into
the light curve, again with very sharp in and out transitions. 

For observation 40124-01-14-00 (MJD 51468.427, panel b) the behaviour was
different: when the source flux was low the power spectrum showed a type-C
shape ($\sim 8.7$ Hz), while during the two peaks in count-rate, when
it reached values close to those of the first half of the previous
observation, the PDS was of type-B ($\sim 6.4$ Hz). The transitions
were again very sharp. In observation 40122-01-01-00 (MJD 51469.360, panel c)
the source showed the opposite behaviour: at high count rates, the power
spectrum transitioned to type-A ($\sim 7.6$ Hz), while at lower fluxes a
type-B PDS was observed ($\sim 6$ Hz), similar to panel (a). This was
clearly seen before and after the gap in the middle of the
observation.

It is worth remarking that in all cases the transitions involve
type-B QPOs. When the source showed a type-B PDS and underwent a fast
transition to a {\it lower} count rate, the PDS changed to type-C;
when on the other hand the transition was to a {\it higher} count
rate, the PDS changed to type-A. In the same way, fast transitions
from type-A to type-B and from type-C to type-B always involved a
decrease and increase in count rate, respectively. Direct transitions
between types A and C were not observed.  Panel (d) of Figure
\ref{trans} (Obs. Id. 40124-01-24-00, MJD 51474.429), shows a fourth type of
rapid transition, which occurred when the source showed a B-Cathedral type
PDS. Corresponding to the dips in the light curve, the PDS changed
its shape, with the two peaks partially losing their coherence while
the red noise increased. Unfortunately, the time intervals in which
this happened were too short for a detailed power-spectrum analysis. 

\begin{figure}[t]
     \centering
     \includegraphics[width=8.5cm]{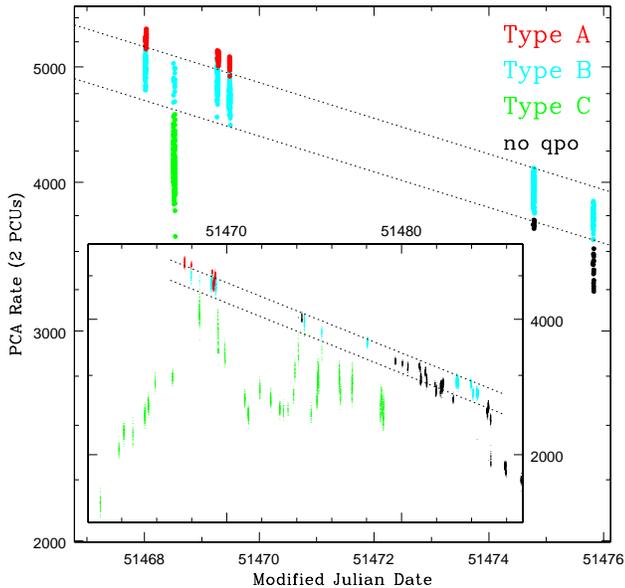}
     \caption{Light curve of the observations showing transitions
     between two different QPO types. The two parallel lines have been drawn
     so as to intersect the transition points. In the inset the lines have
     been extended to the part of the outburst where QPOs
     appear. Different grayscales indicate different QPO types. (A color
     image is available on line) Black
     points correspond to data in which no QPOs were found.}
     \label{expo}
   \end{figure}

In Figure \ref{expo} we plot a light curve of all observations in
which fast transitions were observed. From this figure it is evident
that the count rate at which the transitions occur follow an exponential
trend. Type-B QPOs appear in a narrow count rate range, as can
be seen in the inset of the same figure, where we show the entire
portion of the outburst where low frequency QPOs were detected.

Similar fast transitions between different types of QPO and broad-band noise
components were reported with Ginga from \object{GS~1124--68}
(\cite{Takizawaetal97}) and \object{GX~339--4} itself
(\cite{Miyamotoetal91}). Interestingly, also in these cases the sharp QPO has
a frequency around $\sim$6 Hz.

   \begin{figure}[t]
     \centering
     \includegraphics[width=8.0cm]{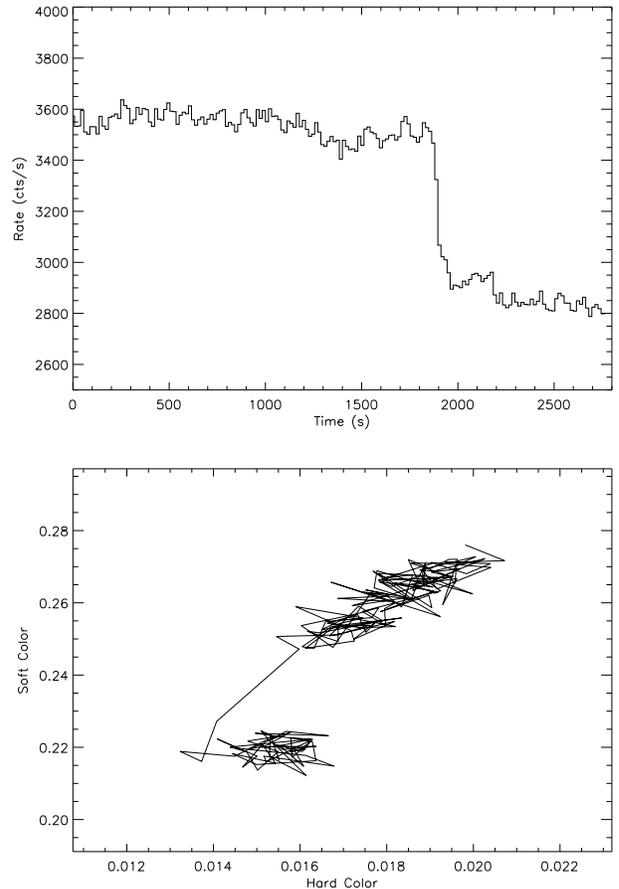}
     \caption{Light curve and color-color diagram for observation
     40124-01-38-01. Only three PCUs were on (MJD: 51485.075).}
     \label{jump}
   \end{figure}

Finally, in Figure \ref{jump} (Obs. Id. 40124-01-38-01), we show a
peculiar event: the source count rate dropped by $\sim$20\% (with an
integrated fractional rms of $\sim$2.5\% before and $\sim$4.0\% after
the drop) in about 100 s. The increase in rms occurs mainly above 5
Hz, with a broad power excess clearly visible around $\sim$6 - 7 Hz
(not shown). The hardness distribution was also different, as can be
seen in the bottom panel of Figure \ref{jump}. During the
observation, the hard and soft colors decreased fairly continuosly,
but simultaneously with the drop in count rate the soft color showed
a clear discontinuity. This sharp transition is also clearly visible 
in Figures \ref{outburst} (MJD: 51485.075) where it can be seen that after the
drop the count rate kept following an approximate exponential trend for
several days, with a slope close to that shown in Figure \ref{zoom}
before the drop, which was thus a conspicuous feature in the overall light
curve of the outburst.


   \begin{figure}[t]
     \centering
     \includegraphics[width=8.5cm]{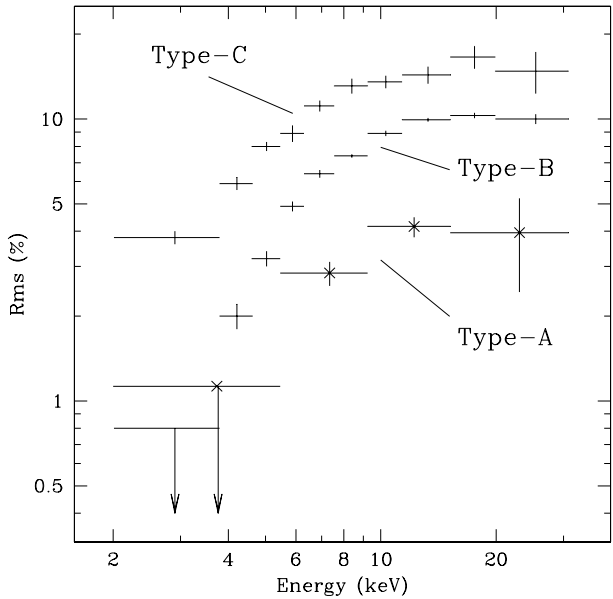}
     \caption{Fractional rms amplitude of the fundamental QPO peak as a
       function of energy for QPO types A (stars, obs: 40124-01-12-00), B
       (40122-01-01-03) and C (40124-01-21-00).}
     \label{rms_ene}
   \end{figure}

\section{Discussion}

We analyzed the RXTE/PCA data from the 1999 outburst of the black
hole candidate \object{XTE~J1859+226}, studying the low frequency QPOs and 
their detailed behaviour. We could classify most of QPOs in three
main types (A, B, C), plus a couple of sub-types, obtaining a
coherent scenario which can be compared to that of other systems such
as \object{XTE~J1550--564}, for which the "ABC" classification was introduced
(\cite{Wijnandsetal99}, \cite{Remillardetal02b}), and \object{GX~339--4}, where
a type-B QPO was found (\cite{Nespolietal03}). We now discuss our
results in terms of relations between derived quantities.

\begin{itemize}

\item {\it Energy dependence:}  In Fig. \ref{rms_ene}, we show the
fractional rms of the QPO as a function of energy for type-A, B and C
spectra. The rms amplitude of the QPOs increases with energy and then
flattens above $\sim$10 keV. The similarity of the trend in the three
QPO types, particularly in the B and C cases, is apparent.
The energy spectra of black hole candidates are often described in
terms of two components, one soft/low energy component and another
hard/high energy component (see e.g. \cite{Tanaka&Lewin95}). 
The observed energy dependence
of the fractional rms clearly leads to the conclusion that the
QPOs are associated with the high-energy component. Notice that
a scenario in which the intrinsic fractional rms of the QPOs (i.e.
the rms normalised only to the source counts from the high energy
component) is constant - with the observed decrease in the rms
towards lower energy being caused by a non-variable soft component -
cannot be ruled out. If this hypothesis were true, the QPOs could
then be interpreted as bare flux oscillations of the high energy
component only, without photon index changes. Nevertheless, in \object{GRS
1915+105} a similar energy dependence for the type-C QPO was observed,
but the analysis of higher-energy HEXTE data showed that the QPO rms
decreases above 20 keV (\cite{TomsickKaaret01}). Rodriguez et al.
(2004) found however that this cut-off was not always present, and
proposed an explaination in terms of a contribution from the jets to
the hard X-ray component. If this contribution is significant, and
the QPOs are unrelated to the jet, its presence would thus produce
the observed cut-off in the energy dependence of the QPO rms.

\item {\it Time lags:} In Figure \ref{rms_lags}, we show the
behaviour of the time lags of the fundamental QPO peak as a function
of the total rms of the PDS. A separation among the three different
QPO types is evident: type-C QPOs show a strong and well defined
trend, with smaller soft lags for larger rms, while type-B QPOs have
hard lags and are clustered around rms of a few percent. The latter
show also evidence for a dependence of the time-lags on the rms.
Finally, type-A lags are negative and show  large scatter.  The
separation of the three types is evident also in Figure \ref{lags_fund_harm}
(left panel), where we plot the time lags of the fundamental and of
the second harmonic. Type-B and type-C QPOs show a similar behaviour:
the lags of the fundamental and those of the second harmonic, with the
exeption of a few cases with large error bars, always have 
opposite signs, resulting in a clustering of the points
in the second and fourth quadrant. This analysis could not be extended to
type-A QPOs, since they don't show any harmonic peak.

   \begin{figure}[t]
     \centering
     \includegraphics[width=8.5cm]{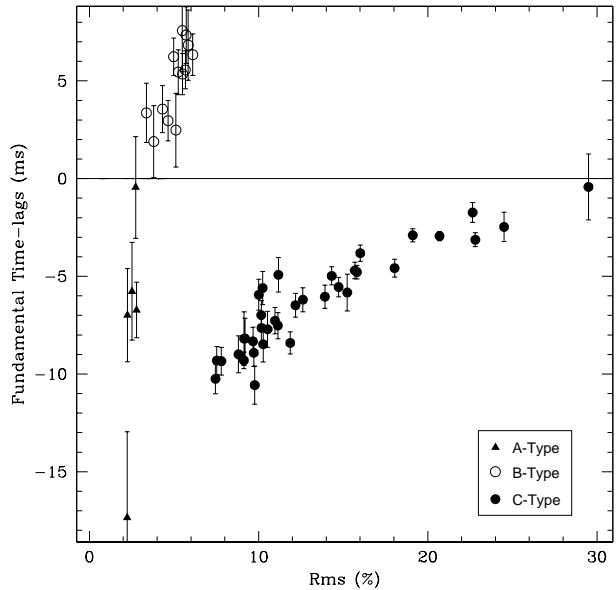}
     \caption{Total fractional rms (0.03-64 Hz) vs. time-lags of the 
fundamental QPO peak.}
     \label{rms_lags}
   \end{figure}

   \begin{figure*}[t]
     \centering
     \includegraphics[width=16.5cm]{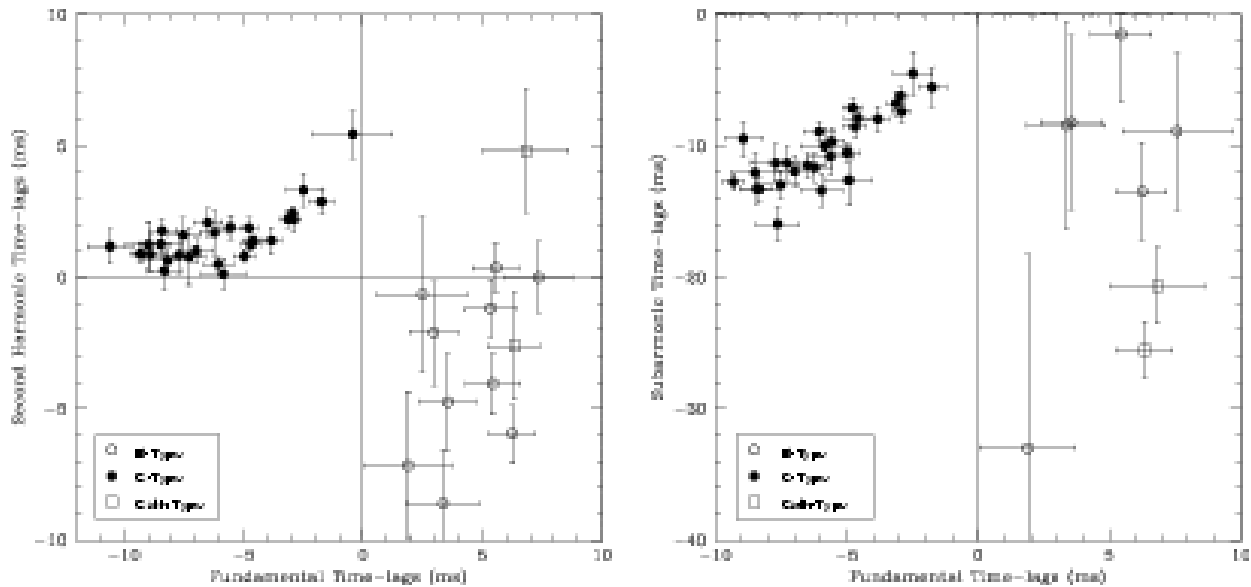}
     \caption{Time-lags of the fundamental QPO peak vs. time-lags of its second
     harmonic ({\it left panel}) and subharmonic ({\it right panel}).}
     \label{lags_fund_harm}
   \end{figure*}

This peculiar pattern in the time lags of the different harmonic components was
already noticed by Remillard et al. (2002b) in the PDS of \object{XTE
J1550--564}. Different lags in these components are difficult to interpret. It
is of course possible that the oscillations at different harmonics have a
different physical origin, while connected to the same underlying `clock', in
which case the difference in sign of time lags would be naturally associated
to these different physical mechanisms. However, if one assumes the more
natural scenario in which the harmonic content of the QPO is simply the result
of the non-sinuoidal shape of the oscillation, the higher harmonics do no have
a physical meaning by themselves (for a discussion of the subharmonic see
below). Such strong differences in time lags would then indicate that the
shape of the oscillation is different at different energies, in a way which is
highly reproducible. In order to understand this phenomenon in detail, a
theoretical model for the shape of the oscillation is needed.

\item {\it Subharmonic peaks:} The right panel of Figure
\ref{lags_fund_harm}, giving the time lags of the subharmonic versus
those of the fundamental, shows that subharmonic lags are {\it
always} negative. This behaviour is independent of the QPO type and
thus of the sign of time lags at the fundamental and second harmonic
peak frequencies. This somehow makes the subharmonic oscillations
stand out, and suggests a common origin for it in both B and C QPO
types.

The existence of a subharmonic peak opens then a serious issue about the
physical mechanism that would produce them. Even though there are
other astrophysical examples of subharmonics (see
e.g. \cite{Aikawa&Antonello00} for the case of Cepheids and
\cite{Masser&Tagger97} for the case of disk galaxies), there are still no
generally-accepted physical explainations. A basic mathematical
explaination involves a subharmonic resonance, which can arise in
forced non-linear oscillators. It can be shown (see for example
\cite{Butikov02}) that for such an oscillator a subharmonic resonance
of order $n$ can occur when the driving frequency is close to an
integer multiple $n$ of the natural fundamental frequency. In this
scenario, the frequency of $\sim$6 Hz seems to play an important
role: from Figure \ref{subh} it is clear that the coherence of the type-B
subharmonic peak is higher when the fundamental peak is close to 6 Hz. It is
worth noticing however that type-C QPOs do not show any peculiar behavior when
at this frequency. In a
forthcoming paper we shall discuss in greater detail the role of this
frequency in different classes of sources, BHCs (see
e.g. \cite{Nespolietal03}), Z-sources and atoll-sources (see e.g. \cite{1820}).

There is of course an alternative scenario in which the frequency of the
subharmonic peak is in reality the fundamental frequency of the system. This
hypothesis would find some support in the fact that for all types of QPOs the
subharmonic always shows the same sign of lags. If this is the case,
there would be no need for an explanation of subharmonic peaks; however, the
subharmonic is not always observed and almost never (with the exception of the
`cathedral' cases) the strongst peak. Moreover, any model under this scenario
would have to explain the absence of the third harmonic.

   \begin{figure*}[t]
     \centering
     \includegraphics[width=18cm]{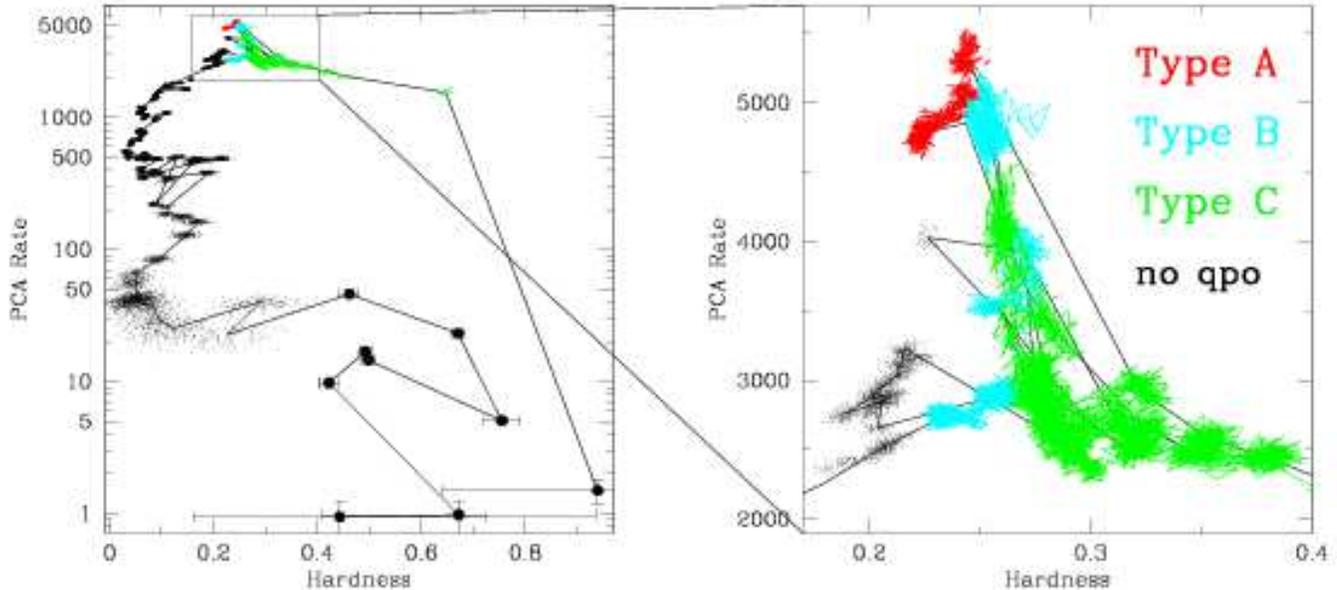}
     \caption{{\it Left panel:} Hardness-Intensity diagram (see Figure 1 for
     variables definition). The time resolution is 16 s except for the first
     and the last six observations, which have one point for each observation
     in order to improve the statistics. Different grayscales indicate
     different QPO types. (A color image is available on line) {\it
     Right panel:} enlargement of the region of the diagram where most of type
     A-B-C QPOs appear.}
     \label{int_col}
   \end{figure*}

\item {\it Color evolution:} In Figure \ref{int_col} we
plot the Hardness-Intensity diagram of the outburst, which shows
clear analogies with diagrams of other BHCs (see \cite{Homanetal01},
\cite{Belloni03}, \cite{Rossietal03}). The energy spectrum is hard at
the beginning of the outburst, quickly softening as the count rate
approaches the outburst peak. Here, the flux shows strong variations,
hardening and softening on short time scales. It is during this phase, at the
highest count rates, that the QPOs were nearly always present. In particular,
from the right panel of Figure \ref{int_col} it can be seen that type-A QPOs
appear when the hardness indicates that the energy spectrum is {\it softer}
than when type-C QPOs are present, while type-B QPOs appear at hardness values
overlapping those of type-C and type-A QPOs, but always restricted
between 0.2-0.3. A similar correlation between QPO types and hardness was
found in \object{XTE~J1550--564} (\cite{Homanetal01}). 
From the result reported in Section 4, it is thus clear that
the type-B QPO can be associated with spectral transitions. This result
is also found in \object{GX~339--4} (Belloni et al. 2004, in prep.). Either
this oscillation is excited in a small range of disk parameters (accretion
rate?), or its frequency depends only weakly on these parameters. In either
cases, this oscillation seems to be an important ingredient in the evolution
of the outbursts of a number of BHT.

After this phase, the count rate first decreased while
the spectrum still softens, then it showed a hard flaring episode,
after which it continued to decrease while maintaining roughly a
constant hardness. Finally, after more than four months from the
beginning of the outburst, the energy spectrum hardened, moving
towards hardness values comparable to those of the first
observation.
Following Homan \& Belloni (in prep.), the roughly horizontal
branches at the top and at the bottom of Fig. \ref{int_col} would
therefore correspond to the very high/intermediate state 
(see \cite{Homanetal01}), appearing
at different flux values, while the vertical branch to the left would
represent the high/soft State (as can be inferred also from the low
rms values). Since the early part of the rise of the outburst has not been
observed we can only argue, in
analogy with other sources, that the source moved along a roughly
vertical branch on the right, which would correspond to the hard
state.

\end{itemize}

\section{Conclusions}

From the large set of observations presented here, the classification of 
low-frequency QPOs into three types emerges strenghtened and is extended to 
another source, \object{XTE~J1859+226}.  In particular, the type B and C QPOs
seem to be a key ingredient that is found in a number of sources. Their
different properties (while at the same centroid frequency of $\sim$6 Hz)
might provide clues to the understanding of the physical mechanisms during the
VHS/IS of BHCs and are probably related to the two ``flavors'' of VHS/IS
observed in these systems (see e.g. \cite{Miyamotoetal93}). The observed energy
  dependence, that rules out a direct disk origin, is an important constrain
  for the future physical identification of these features. The difference in
  time-lag
behaviour of the different peaks and the presence of the subharmonic peak are
challenging features of the QPO phenomenon. In particular our results provide
additional evidence that the frequency of $\sim$6 Hz is related to some
fundamental process. The evidence of a treshold triggering the $\sim$6 Hz QPO,
superimposed to the roughly exponential decay of the outburst, furthermore
suggests the existence of a second parameter in addition to the mass accretion
rate probably responsible for the long time-scale evolution of the source.

In addition, after the analysis of yet another transient source, it
is clear that the outbursts of these systems, in their general
evolution, show strong similarities. This should be examined in the
light of theoretical models for the high-energy emission and possibly
even used as another identification tool for these sources.

\begin{acknowledgements}

This work was partially supported by MIUR under CO-FIN grant 2002027145.
Jeroen Homan acknowledges support from NASA.

\end{acknowledgements}




\begin{thebibliography}{}

\bibitem[Aikawa \& Antonello 2000]{Aikawa&Antonello00}
Aikawa, T., \& Antonello, E., 2000, A\&A, 363, 593

\bibitem[Belloni \& Hasinger 1990]{Belloni90}
Belloni, T., \& Hasinger, G., 1990, A\&A, 230, 103

\bibitem[Belloni et al. 1997]{Bellonietal97}
Belloni, T., van der Klis, M., Lewin, W. H. G.,

\bibitem[Belloni et al. 2002]{Bellonietal02}
Belloni, T., Psaltis, D., \& van der Klis, M., 2002, ApJ, 572,392

\bibitem[Belloni 2003]{Belloni03}
Belloni, T. 2003, to appear in Proc. of the II BeppoSAX Meeting: "The Restless
High-Energy Universe" (Amsterdam, May 5-8, 2003), E.P.J. van den Heuvel,
J.J.M. in 't Zand, and R.A.M.J. Wijers Eds

\bibitem[Belloni, Parolin \& Casella 2004]{1820}
Belloni, T., Parolin, I., \& Casella, P., 2004, A\&A, accepted

\bibitem[Brocksopp et al. 2002]{Brocksoppetal02}
Brocksopp, C., Fender, R. P., McCollough, M., Pooley, G. G., Rupen, M. P.,
Hjellming, R. M., de la Force, C. J., Spencer, R. E., Muxlow, T. W. B.,
Garrington, S. T., Trushkin, S., 2002, MNRAS, 331, 765

\bibitem[Butikov 2002]{Butikov02}
Butikov, E. I., 2002, J. Phys. A: Math. Gen., 35, 6209

\bibitem[Cui 1998]{Cui98}
Cui, W., 1998, in ``High Energy Processes in Accreting Black Holes'', eds
Poutanen J. \& Svensson R., (Graftavallen, Sweden), June 1998
(astro-ph/9809408)

\bibitem[Cui et al. 1999]{Cuietal99}
Cui, W., Zhang, S. N., Chen, W., \& Morgan, E. H., 1999, ApJ, 512, 43

\bibitem[Cui et al. 2000]{Cuietal00}
Cui, W., Shrader, C. R., Haswell, C. A., \& Hynes, R. I., 2000, ApJ, 535, L123

\bibitem[Dieters et al. 2000]{Dietersetal00}
Dieters, S. W., Belloni, T., Kuulkers, E., Woods, P., Cui, W., Zhang, S. N.,
Chen, W., van der Klis, M., van Paradijs, J., Swank, J., Lewin, W. H. G., \&
Kouveliotou, C., 2000, ApJ, 538, 307

\bibitem[Focke et al. 2000]{Fockeetal00}
Focke, W. B., Markwardt, C. B., Swank, J. H., \& Taam, R. E., 2000, in
"Rossi2000: Astrophysics with the Rossi X-ray Timing Explorer",
Greenbelt, March 2000, E104

\bibitem[Garnavich, Stanek, \& Berlind 1999]{Garnavichetal99}
Garnavich, P. M., Stanek, K. Z., \& Berlind, P., 1999, IAU Circ. 7276

\bibitem[Garnavich \& Quinn 2000]{Garnavichetal00}
Garnavich, P. M., \& Quinn, J., 2000, IAU Circ. 7388

\bibitem[Homan et al. 2001]{Homanetal01}
Homan, J., Wijnands, R., van der Klis, M., Belloni, T., van Paradijs, J.,
Klein-Wolt, M., Fender, R., \& Mendez, M., 2001, ApJS, 132, 377

\bibitem[Markwardt et al. 1999]{Markwardtetal99}
Markwardt, C. B., Marshall, F. E., \& Swank, J. H., 1999, IAUC 7274

\bibitem[Markwardt et al. 2001]{Markwardtetal01}
Markwardt, C. B., Focke, W. B., \& Swank, J. H., 2001, in ``X-ray Emission
from Accretion onto Black Holes'' Eds. T. Yaqoob and J. H. Krolik, Johns
Hopkins University, Baltimore, published electronically

\bibitem[Masser \& Tagger 1997]{Masser&Tagger97}
Masser, F., \& Tagger, M., 1997, A\&A, 322, 442

\bibitem[McClintock \& Remillard 2004]{McC&R04}
McClintock, J. E., \& Remillard, R. A., 2004, to appear in "Compact Stellar
X-ray Sources", eds. W.H.G. Lewin and M. van der Klis, Cambridge University
Press, Cambridge

\bibitem[Mendez \& van der Klis 1997]{Mendez&vdK97}
Mendez, M., \& van der Klis, M., 1997, ApJ, 479, 926

\bibitem[Mendez et al. 1998]{Mendezetal98}
Mendez, M., Belloni, T., \& van der Klis, M., 1998, ApJ, 499, L187

\bibitem[Miyamoto et al. 1991]{Miyamotoetal91}
Miyamoto, S., Kimura, K., Kitamoto, S., Dotani, T., \& Ebisawa, K., 1991, ApJ,
383, 784

\bibitem[Miyamoto et al. 1993]{Miyamotoetal93}
Miyamoto, S., Iga, S., Kitamoto, S., \& Kamado, Y., 1993, ApJ, 403, L39

\bibitem[Nespoli et al. 2003]{Nespolietal03}
Nespoli, E., Belloni, T., Homan, J., Miller, J. M., Lewin, W. H. G., Mendez,
M., \& van der Klis, M., 2003, A\&A, 412, 235

\bibitem[Park et al. 2003]{Parketal03}
Park, S. Q., Miller, J. M., McClintock, J. E., Remillard, R. A., Orosz, J. A.,
Shrader, C. R., Hunstead, R. W., Campbell-Wilson, D., Ishwara-Chandra, C. H.,
Rao, A. P., \& Rupen, M. P., 2003, ApJ, submitted

\bibitem[Pooley \& Hjellming 1999]{PooleyHjellming99}
Pooley, G. G., \& Hjellming, R. M., 1999, IAU Circ. 7278

\bibitem[Reig et al. 2000]{Reigetal00}
Reig, P., Belloni, T., van der Klis, M., Mendez, M., Kylafis, N. D., \& Ford,
E. C., 2000, ApJ, 541, 883

\bibitem[Remillard et al. 1999]{Remillardetal99}
Remillard, R. A., Morgan, E. H., McClintock, J. E., Bailyn, C. D., \& Orosz,
J. A., 1999, ApJ, 522, 397

\bibitem[Remillard et al. 2002a]{Remillardetal02a}
Remillard, R. A., Muno, M. P., McClintock, J. E., \& Orosz, J. A., 2002a, in
New Views on Microquasars, 49

\bibitem[Remillard et al. 2002b]{Remillardetal02b}
Remillard, R. A., Sobczak, G. J., Muno, M. P., \& McClintock, J. E., 2002b,
ApJ, 564, 962

\bibitem[Rodriguez et al. 2004]{Rodriguezetal04}
Rodriguez, J., Fuchs, Y., Hannaikainen, D., Vilhu, O., Shaw, S.,
  Belloni, T., \& Corbel, S., 2004, Proc. of the 5h INTEGRAL workshop,
  (Munich, Feb 16-20, 2004), to be published by EDP

\bibitem[Rossi et al. 2003]{Rossietal03}
Rossi, S., Homan, J., Miller, J. M., \& Belloni, T. 2003, to appear in
Proc. of the II BeppoSAX Meeting: "The Restless High-Energy Universe"
(Amsterdam, May 5-8, 2003), E.P.J. van den Heuvel, J.J.M. in 't Zand, and
R.A.M.J. Wijers Eds

\bibitem[Sobczak et al. 2000]{Sobczaketal00}
Sobczak, G. J., McClintock, J. E., Remillard, R. A., Cui, W., Levine, A. M.,
Morgan, E. H., Orosz, J. A., \& Bailyn, C. D., 2000, ApJ, 531, 537

\bibitem[Takizawa et al. 1997]{Takizawaetal97}
Takizawa, M., Dotani, T., Mitsuda, K., Matsuba, E., Ogawa, M., Aoki, T., Asai,
K., Ebisawa, K., Makishima, K., Miyamoto, S., Iga, S., Vaughan, B., Rutledge,
R. E., \& Lewin, W. H. G., 1997, ApJ, 489, 272

\bibitem[Tanaka \& Lewin 1995]{Tanaka&Lewin95}
Tanaka, Y., \& Lewin, W. H. G., 1995, in "X-ray binaries", eds. Lewin,
W. H. G., van Paradijs, J. and van den Heuvel, E. P. J., Cambridge University
Press, Cambridge, p. 126

\bibitem[Tomsick \& Kaaret 2000]{TomsickKaaret00}
Tomsick, J. A., \& Kaaret, P., 2000, ApJ, 537, 448

\bibitem[Tomsick \& Kaaret 2001]{TomsickKaaret01}
Tomsick, J. A., \& Kaaret, P., 2001, ApJ, 548, 401

\bibitem[van der Klis 1995]{vanderKlis95}
van der Klis, M., 1995, in "X-ray binaries", eds. Lewin, W. H. G., van
Paradijs, J. and van den Heuvel, E. P. J., Cambridge University Press,
Cambridge, p. 252

\bibitem[van der Klis 2000]{vanderKlis00}
van der Klis, M., 2000, ARAA, 38, 717

\bibitem[van der Klis 2004]{vanderKlis04}
van der Klis, M.,  2004, to appear in "Compact Stellar
X-ray Sources", eds. W.H.G. Lewin and M. van der Klis, Cambridge University
Press, Cambridge

\bibitem[Zhang 1995]{Zhang95}
Zhang, W., 1995, XTE/PCA Internal Memo

\bibitem[Zhang et al. 1995]{Zhangetal95}
Zhang, W., Jahoda, K., Swank, J. H., Morgan, E. H., \& Giles. A. B., 1995,
ApJ, 449, 930

\bibitem[Zurita et al. 2002]{Zuritaetal02}
Zurita, C., Sánchez-Fernández, C., Casares, J., Charles, P. A., Abbott, T. M.,
Hakala, P., Rodríguez-Gil, P., Bernabei, S., Piccioni, A., Guarnieri, A.,
Bartolini, C., Masetti, N., Shahbaz, T., Castro-Tirado, A., \& Henden, A.,
2002, MNRAS, 334, 999

\bibitem[Wijnands et al. 1999]{Wijnandsetal99}
Wijnands, R., Homan, J., \& van der Klis, M., 1999, ApJ, 526, L33

\bibitem[Wijnands et al. 2001]{Wijnandsetal01}
Wijnands, R., Mendez, M., Miller, J. M., \& Homan, J., 2001, MNRAS, 328, 451

\bibitem[Wood et al. 1999]{Woodetal99}
Wood, A., Smith, D. A., Marshall, F. E., Swank, J., 1999, IAUC 7274

\end{thebibliography}
\end{document}